\documentclass[sigconf]{acmart}

\usepackage{bigstrut}
\usepackage{booktabs}
\usepackage{multirow}
\usepackage{microtype}
\usepackage{graphicx}
\usepackage{subfig}
\usepackage{overpic}
\usepackage{balance}
\usepackage{amsthm}
\usepackage{array}
\usepackage{clrscode}
\usepackage{multicol}
\usepackage{float}
\usepackage{newfloat}
\usepackage{color}
\usepackage{xcolor}
\usepackage{amsopn}
\usepackage{mathrsfs}
\usepackage{mathtools}
\usepackage{amsmath}
\usepackage{arydshln}
\usepackage{blkarray}
\usepackage{enumerate}
\usepackage{courier}
\usepackage{rotating}
\usepackage{bm}
\usepackage{wrapfig}
\usepackage{ragged2e}
\usepackage[misc]{ifsym}
\usepackage{threeparttable}
\usepackage{makecell}
\usepackage{adjustbox} % For adjusting table size
\usepackage{enumitem}
\usepackage{hyperref}
\usepackage{url}
\usepackage{xspace}
\usepackage{comment}
\usepackage{changepage}
\usepackage{algorithm}
\usepackage{algpseudocode} 
\usepackage{algorithmicx} 
\usepackage[ruled, vlined, nofillcomment, linesnumbered, algo2e]{algorithm2e}

\definecolor{codegreen}{rgb}{0,0.3,0.6}
\definecolor{codegray}{rgb}{0.5,0.5,0.5}

\newcommand{\ie}{\emph{i.e.,}\xspace}
\newcommand{\eg}{\emph{e.g.,}\xspace}

\newcommand{\paratitle}[1]{\vspace{1.5ex}\noindent\textbf{#1}}
\newcommand{\wrt}{w.r.t.\xspace}

\newcommand{\ignore}[1]{}

\AtBeginDocument{%
  \providecommand\BibTeX{{%
    \normalfont B\kern-0.5em{\scshape i\kern-0.25em b}\kern-0.8em\TeX}}}

\setcopyright{acmlicensed}
\copyrightyear{2018}
\acmYear{2018}
\acmDOI{XXXXXXX.XXXXXXX}

\acmConference[Conference acronym 'XX]{Make sure to enter the correct
  conference title from your rights confirmation emai}{June 03--05,
  2018}{Woodstock, NY}
\acmISBN{978-1-4503-XXXX-X/18/06}

\begin{document}

\title{Enhancing Graph Contrastive Learning with Reliable and Informative Augmentation for Recommendation}  

\settopmatter{authorsperrow=4}
\author{Bowen Zheng}
\orcid{0009-0002-3010-7899}
\affiliation{%
% Gaoling School of Artificial Intelligence, 
    \institution{
    % Gaoling School of Artificial Intelligence,  
    Renmin University of China}
    \city{Beijing}
    \country{China}
}
\email{bwzheng0324@ruc.edu.cn}

\author{Junjie Zhang}
% \orcid{}
\affiliation{%
% Gaoling School of Artificial Intelligence, 
    \institution{
    % Gaoling School of Artificial Intelligence,  
    Renmin University of China}
    \city{Beijing}
    \country{China}
}
\email{junjie.zhang@ruc.edu.cn}

\author{Hongyu Lu}
% \orcid{}
\affiliation{%
% Gaoling School of Artificial Intelligence, 
    \institution{WeChat, Tencent}
    \city{Guangzhou}
    \country{China}
}
\email{luhy94@gmail.com}

\author{Yu Chen}
% \orcid{}
\affiliation{%
% Gaoling School of Artificial Intelligence, 
    \institution{WeChat, Tencent}
    \city{Beijing}
    \country{China}
}
\email{nealcui@tencent.com}

\author{Ming Chen}
% \orcid{}
\affiliation{%
% Gaoling School of Artificial Intelligence, 
    \institution{WeChat, Tencent}
    \city{Guangzhou}
    \country{China}
}
\email{mingchen@tencent.com}

\author{Wayne Xin Zhao
%$\dagger$
\textsuperscript{\Letter}
}
\orcid{0000-0002-8333-6196}
\affiliation{
    \institution{
    % Gaoling School of Artificial Intelligence,  
    Renmin University of China}
    \city{Beijing}
    \country{China}
}
\email{batmanfly@gmail.com}

\author{Ji-Rong Wen}
\orcid{0000-0002-9777-9676}
\affiliation{
    \institution{
    % Gaoling School of Artificial Intelligence,  
    Renmin University of China}
    \city{Beijing}
    \country{China}
}
\email{jrwen@ruc.edu.cn}

\thanks{\Letter \ Corresponding author.}
% \thanks{$\dagger$ Beijing Key Laboratory of Big Data Management and Analysis Methods.}

\renewcommand{\shortauthors}{Bowen Zheng, et al.}

\begin{abstract}
Graph neural network~(GNN) has been a powerful approach in collaborative filtering~(CF) due to its ability to model high-order user-item relationships.
Recently, to alleviate the data sparsity and enhance representation learning, many efforts have been conducted to integrate contrastive learning~(CL) with GNNs. 
Despite the promising improvements, the contrastive view generation based on structure and representation perturbations in existing methods potentially disrupts the collaborative information in contrastive views, resulting in limited effectiveness of positive alignment.

To overcome this issue, we propose CoGCL, a novel framework that aims to enhance graph contrastive learning by constructing contrastive views with stronger collaborative information via discrete codes.
The core idea is to map users and items into discrete codes rich in collaborative information for reliable and informative contrastive view generation.
To this end, we initially introduce a multi-level vector quantizer in an end-to-end manner to quantize user and item representations into discrete codes. 
Based on these discrete codes, we enhance the collaborative information of contrastive views by considering neighborhood structure and semantic relevance respectively.
For neighborhood structure, we propose virtual neighbor augmentation by treating discrete codes as virtual neighbors, which expands an observed user-item interaction into multiple edges involving discrete codes.
Regarding semantic relevance, we identify similar users/items based on shared discrete codes and interaction targets to generate the semantically relevant view.
Through these strategies, we construct contrastive views with stronger collaborative information and develop a triple-view graph contrastive learning approach.
Extensive experiments on four public datasets demonstrate the effectiveness of our proposed approach. Moreover, detailed analyses highlight our contribution in enhancing graph CL for recommendation.
Our code is available at \textcolor{blue}{\url{https://github.com/RUCAIBox/CoGCL}}

\end{abstract}

\begin{CCSXML}
<ccs2012>
   <concept>
       <concept_id>10002951.10003317.10003347.10003350</concept_id>
       <concept_desc>Information systems~Recommender systems</concept_desc>
       <concept_significance>500</concept_significance>
    </concept>
 </ccs2012>
\end{CCSXML}

\ccsdesc[500]{Information systems~Recommender systems}

\keywords{Recommendation, Collaborative Filtering, Graph Contrastive Learning}

\maketitle

\section{Introduction}
\label{sec:intro}

In the literature of recommender systems, collaborative filtering~(CF) based on graph neural network~(GNN) has showcased significant success in recommendation systems due to its ability to model high-order user-item relationships~\cite{DBLP:conf/sigir/Wang0WFC19,DBLP:conf/sigir/0001DWLZ020,DBLP:journals/tors/GaoZLLQPQCJHL23}.
This approach typically involves organizing user-item interaction data into a bipartite graph and learning node representations that contain collaborative knowledge from the graph structure.
However, given the sparsity of user behaviors, GNN-based methods often struggle with limited graph edges and insufficient supervision signals. 
This challenge hinders the ability to develop high-quality user and item representations~\cite{DBLP:journals/csur/WuSZXC23,DBLP:conf/sigir/WuWF0CLX21,DBLP:conf/kdd/WangYM000M22}, which are vital for improving recommendation.
To address this challenge, recent studies propose to integrate contrastive learning~(CL)~\cite{DBLP:conf/icml/ChenK0H20,DBLP:conf/emnlp/GaoYC21,DBLP:journals/corr/abs-2011-00362} with GNN-based CF to incorporate self-supervised signals. 

According to how the contrastive views are constructed, existing Graph CL-based methods can be divided into two categories: structure augmentation and representation augmentation.
Structure augmentation perturbs the graph structure to create augmented graphs, which are subsequently used by the GNN to generate contrastive node representations~\cite{DBLP:conf/sigir/WuWF0CLX21,DBLP:conf/iclr/Cai0XR23,DBLP:conf/sigir/LiXRY0023,DBLP:conf/sigir/RenXZY023}.
As a representative method, SGL~\cite{DBLP:conf/sigir/WuWF0CLX21} adopts stochastic node/edge dropout to construct augmented graphs as contrastive views.
Representation augmentation involves encoding additional representations of nodes from the interaction graph for CL~\cite{DBLP:conf/sigir/LeeKJPY21,DBLP:conf/www/LinTHZ22,DBLP:conf/sigir/XiaHXZYH22,DBLP:conf/sigir/YuY00CN22,DBLP:journals/tkde/YuXCCHY24,DBLP:conf/kdd/TangDSCXYHJL24}.
Particularly, SimGCL~\cite{DBLP:conf/sigir/YuY00CN22} perturbs the node embedding by adding random noise to generate contrastive views.
Despite their effectiveness, existing approaches still suffer from unexpected self-supervised signals~\cite{DBLP:conf/iclr/Cai0XR23,DBLP:conf/sigir/LiXRY0023}. 
Contrastive view generation based on perturbations potentially \emph{disrupt collaborative information} within contrastive views. 
More precisely, in recommendation scenarios where user behaviors are scarce, structural perturbations may lose key interactions of sparse users~\cite{DBLP:conf/sigir/LiXRY0023,DBLP:conf/sigir/YuY00CN22,DBLP:conf/iclr/Cai0XR23}.
And the random noise added to node embeddings may interfere with the implicit collaborative semantics in node representations~\cite{DBLP:conf/sigir/YangWW0HZZW23,DBLP:conf/iclr/Cai0XR23}.
In addition, the empirical analysis in Section ~\ref{sec:emp_ana} confirms that the \emph{alignment} between positive pairs based on perturbations is not as effective as expected, and the model performance significantly relies on the representation \emph{uniformity} across different instances facilitated by CL.
% as information propagates across the graph.

\begin{figure}[]
\centering
\includegraphics[width=0.98\linewidth]{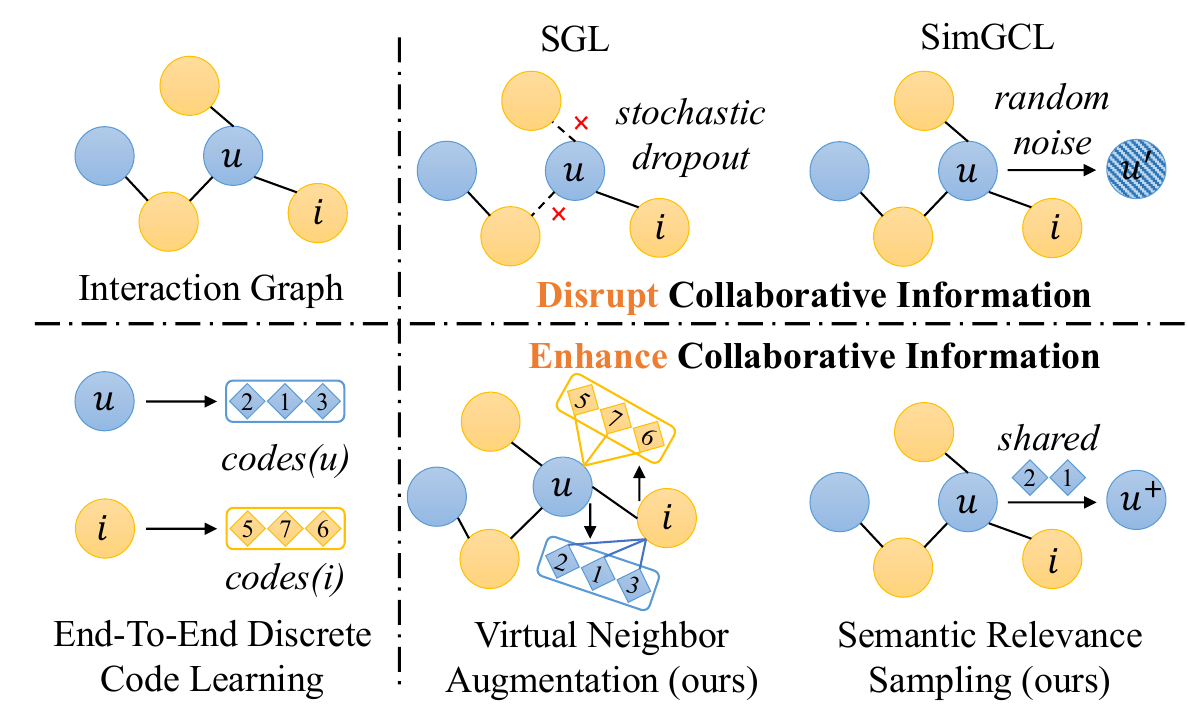}
\caption{Comparison of current graph CL-based methods (\eg SGL~\cite{DBLP:conf/sigir/WuWF0CLX21}, SimGCL~\cite{DBLP:conf/sigir/YuY00CN22}) that disrupt collaborative information within contrastive views and the proposed approach that enhances collaborative information.}
\label{fig:intro}
\end{figure}

Considering these issues, we aim to construct higher-quality contrastive views to \emph{enhance collaborative information}.
Specifically, we strive to maintain both \emph{reliability} and \emph{informativeness} for contrastive view generation.
For reliability, we anticipate that the structural information introduced by graph augmentation is well-founded rather than arbitrary, that is, based on the observed user-item interactions.
% For reliability, we expect the structural information introduced by graph augmentation to be well-founded rather than arbitrary
% rather than stochastic.
% To this end, 
Our idea is to represent each user or item as a tuple of discrete IDs (called \emph{codes} in this paper) associated with collaborative information.
% To this end, our idea is to introduce discrete IDs (called \emph{codes} in this paper) associated with collaborative information to represent each user or item.
Given the user and item codes, as shown in Figure~\ref{fig:intro}, we can naturally expand a ``\texttt{u-i}'' interaction edge to several ``\texttt{u-codes(i)}'' and ``\texttt{codes(u)-i}'' edges.
% , which is reliable as it 
For informativeness, this code-based augmentation can enhance neighborhood structure and effectively alleviate the sparsity of the interaction graph by treating the codes as virtual neighbors.
Furthermore, sharing discrete codes between different users/items indicates their relevance of collaborative semantics, such as $u$ and $u^+$ in Figure~\ref{fig:intro}.
To develop our methodology, we focused on (a) how to elegantly learn discrete codes associated with rich collaborative information and (b) how to integrate the learned discrete codes into the graph CL framework to improve recommendation.

In this paper, we propose CoGCL, a reliable and informative graph CL approach aiming to construct contrastive views that imply stronger collaborative information by introducing discrete codes.
To map users and items into discrete codes rich in collaborative information, we learn a multi-level vector quantizer in an end-to-end manner to quantize user and item representations encoded by GNN into discrete codes.
Subsequently, the learned discrete codes are adopted to enhance the collaborative information of contrastive views in two aspects: \emph{neighborhood structure} and \emph{semantic relevance}.
For neighborhood structure, we conduct virtual neighbor augmentation by treating discrete codes as virtual neighbors based on existing interactions. 
This process serves to enhance the node's neighbor information and alleviate interaction sparsity in contrasting views.
For semantic relevance, we identify users/items that share discrete codes or interaction targets as semantically similar for positive sampling.
By aligning users/items with semantic relevance via CL, we can further enhance the integration of collaborative semantics.
Through the above strategies, we can generate various contrastive views with stronger collaborative information.
Finally, a triple-view graph contrastive learning approach is proposed to achieve alignment across the augmented nodes and similar users/items.
The contributions in this paper can be summarized as follows:

$\bullet$ We present a reliable and informative graph CL approach, namely CoGCL, which constructs contrastive views that imply stronger collaborative information via discrete codes.

$\bullet$ We propose an end-to-end method to elegantly learn discrete codes for users and items. These discrete codes are employed to enhance the collaborative information of contrastive views in terms of both neighborhood structure and semantic relevance.

$\bullet$ Extensive experiments on four public datasets show that our approach consistently outperforms baseline models. Further in-depth analyses illustrate the crucial role that our designed components play in enhancing graph CL for recommendation.

% The contributions in this paper can be summarized as follows:
% \begin{itemize}
% \item We empirically demonstrate the contributions of local clustering and global uniformity in existing graph CL-based methods, respectively, and find that local clustering on existing contrastive views is ineffective.
% \item We present a novel graph CL framework, CoGCL, which improves the efficacy of CL by leveraging discrete codes to construct contrastive views that imply stronger collaborative information.
% \item Extensive experiments on four public datasets show that our approach consistently outperforms baseline models. Further in-depth ablation studies illustrate the crucial role that our designed components play in unleashing the potential of CL.
% \end{itemize}

\section{Preliminary and Empirical Analysis}
\label{sec:preliminary}
In this section, we first overview the common paradigm of graph CL for recommendation.
Subsequently, we conducted a brief empirical analysis to further explore how graph CL works in CL.

\subsection{Graph CL for Recommendation}
Given user and item sets $\mathcal{U}$ and $\mathcal{I}$ respectively, let $\mathbf{R} \in \{0,1\}^{|\mathcal{U}| \times |\mathcal{I}|}$ represent the user-item interaction matrix, where $\mathbf{R}_{u,i} = 1$ if there is an observed interaction between user $u$ and item $i$, otherwise $\mathbf{R}_{u,i} = 0$. 
Based on the interaction data $\mathbf{R}$, GNN-based CF methods construct a bipartite graph $\mathcal{G}=(\mathcal{V}, \mathcal{E})$, where the node set $\mathcal{V}=\{\mathcal{U} \cup \mathcal{I}\}$ includes all users and items, and $\mathcal{E}=\{(u,i)| u \in \mathcal{U}, i \in \mathcal{I}, \mathbf{R}_{u,i}=1\}$ denotes the set of interaction edges.

Typically, GNN-based CF methods~\cite{DBLP:conf/sigir/Wang0WFC19,DBLP:conf/sigir/0001DWLZ020} utilize the neighbor aggregation scheme on $\mathcal{G}$ to obtain informative node representations, which can be formulated as follows:
\begin{align}
    \mathbf{Z}^l = \operatorname{GNN}(\mathbf{Z}^{l-1}, \mathcal{G}), \ \ \ \
    \mathbf{Z} ~ = \operatorname{Readout}([\mathbf{Z}^0, \mathbf{Z}^1, \dots , \mathbf{Z}^L]),
\end{align}
where $L$ denotes the number of GNN layers, and $\mathbf{Z}^l \in \mathbb{R}^{|\mathcal{V}| \times d}$ denotes the node representations at the $l$-th GNN layer, capturing the $l$-hop neighbor information. 
Here, $\mathbf{Z}^0$ is the trainable ID embedding matrix. The readout function $\operatorname{Readout}(\cdot)$ is used to summarize all representations for prediction. 
Then, the predicted score is defined as the similarity between the user and item representations (\eg inner product, $\hat{y}_{ui} = z_u^T z_i$). 
For the recommendation optimization objective, most studies use the pairwise Bayesian Personalized Ranking~(BPR)~\cite{DBLP:conf/uai/RendleFGS09} loss for model training, denoted as $\mathcal{L}_{bpr}$.

In addition, the graph CL-based methods~\cite{DBLP:conf/sigir/WuWF0CLX21,DBLP:conf/iclr/Cai0XR23,DBLP:conf/www/LinTHZ22,DBLP:conf/sigir/YuY00CN22} propose to further improve the recommendation performance by performing contrastive learning between two contrastive views.  
Specifically, given two view representations $\mathbf{z}_v^\prime$ and $\mathbf{z}_v^{\prime \prime}$ of a node (\eg obtained by two augmented graphs~\cite{DBLP:conf/sigir/WuWF0CLX21}), the optimization objective of CL based on InfoNCE~\cite{DBLP:journals/corr/abs-1807-03748} loss is:
\begin{align}
    \mathcal{L}_{cl} = - \log \frac{e^{s(\mathbf{z}_v^\prime,\mathbf{z}_v^{\prime \prime}) / \tau} }{e^{s(\mathbf{z}_v^\prime,\mathbf{z}_v^{\prime \prime}) / \tau} + \sum_{\tilde{v} \in \mathcal{V}_{\text{neg}}    } e^{s(\mathbf{z}_v^\prime,\mathbf{z}_{\tilde{v}}^{\prime \prime}) / \tau} },
    \label{eq:infonce}
\end{align}
where $s(\cdot)$ denotes the cosine similarity function, $\tau$ is the temperature coefficient, $v$ is a user/item, and $\mathcal{V}_{\text{neg}}$ denotes the set of negative samples, such as in-batch negatives. 
Finally, the joint learning scheme of graph CL-based CF is outlined as follows:
\begin{align}
    \mathcal{L} = \mathcal{L}_{bpr} + \mu \mathcal{L}_{cl},
\end{align}
where $\mu$ is a hyperparameter for balance between two objectives.

\begin{figure}[]
\centering
\includegraphics[width=0.95\linewidth]{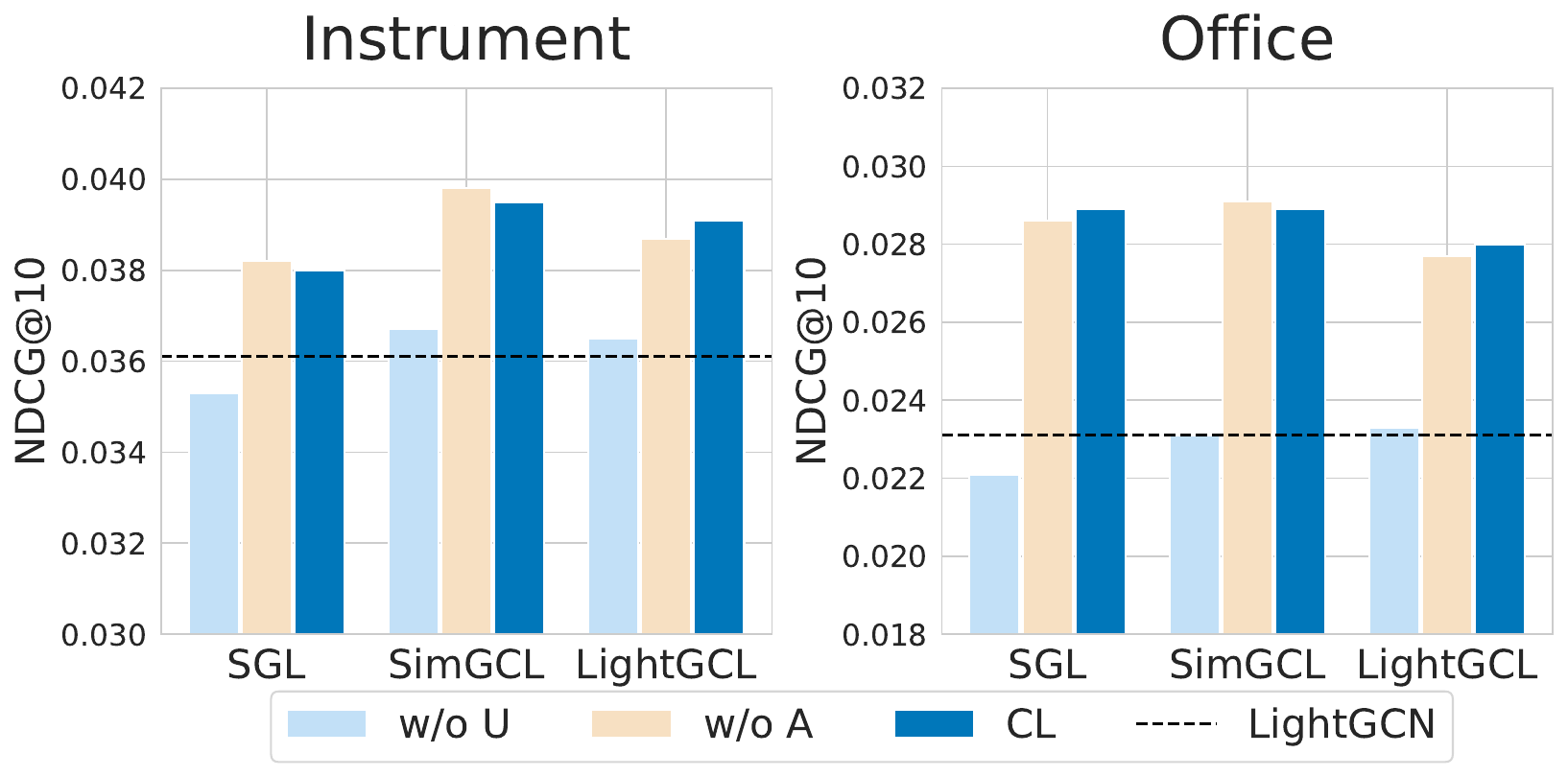}
\caption{Performance comparison of different graph CL-based methods with their variants.}
\label{fig:emp}
\end{figure}

\subsection{Alignment between Perturbed Views is Ineffective}
\label{sec:emp_ana}

% Then, we proceed with an empirical analysis to explore how these two key elements impact the performance of various graph CL models.
To further emphasize our motivation, we proceed with an empirical analysis to explore the limitations of existing methods that disrupt collaborative information.
Following previous works~\cite{DBLP:conf/icml/0001I20,DBLP:conf/emnlp/GaoYC21}, when the number of negative examples is large, the asymptotics of the InfoNCE loss~\eqref{eq:infonce} can be expressed by the following equation:
\begin{align}
    -\frac{1}{\tau} \underset{ (\mathbf{z},\mathbf{z}^{+}) \sim p_{\text{pos}} }{\mathbb{E}}\left[s(\mathbf{z},\mathbf{z}^{+})\right] +\underset{ \mathbf{z} \sim p_{\text{data}}}{\mathbb{E}}\left[\log \underset{ \mathbf{z}^{-} \sim p_{\text{data}}}{\mathbb{E}}\left[e^{s(\mathbf{z},\mathbf{z}^{-}) / \tau}\right]\right],
    \label{eq:asy_infonce}
\end{align}
where $p_{\text{pos}}$ denotes the distribution of positive pairs, and $p_{\text{data}}$ denotes the overall data distribution. Intuitively, the first term maintains the similarity of positive pairs, whereas the second term pushes negative pairs apart. These are formally defined as the \emph{alignment} and \emph{uniformity} of representations on the unit hypersphere~\cite{DBLP:conf/icml/0001I20}.
Here, we try to investigate the contributions of the above two terms by individually disabling their effects. 
Specifically, we conduct experiments on three representative graph CL-based CF models: SGL~\cite{DBLP:conf/sigir/WuWF0CLX21}, SimGCL~\cite{DBLP:conf/sigir/YuY00CN22}, and LightGCL~\cite{DBLP:conf/iclr/Cai0XR23}. 
% Additionally, LightGCN~\cite{DBLP:conf/sigir/0001DWLZ020} serves as a comparison baseline.
For each model, we introduce two variants: 
(a) {\underline{w/o U}} stops the gradient of similarity calculations for negative pairs in Eq.~\eqref{eq:infonce} (using \texttt{detach} function in Pytorch), which leads to the breakdown of uniformity in Eq.~\eqref{eq:asy_infonce}.
(b) {\underline{w/o A}} stops the gradient between positive pairs in Eq.~\eqref{eq:infonce}, resulting in the breakdown of alignment in Eq.~\eqref{eq:asy_infonce}.
From the results in Figure~\ref{fig:emp}, we can observe the following two phenomena:

$\bullet$ Disabling uniformity and only pulling the positive pairs together does not yield a significant improvement compared to LightGCN. Furthermore, \underline{SGL w/o U} produces a decrease in performance.

$\bullet$ Disabling alignment leads to minimal negative impact and might even result in a slight performance improvement.

Generally, alignment between positive examples in the above methods could be ineffective or potentially harmful.
We argue that perturbation methods such as stochastic edge/node dropout (\ie SGL), random noise~(\ie, SimGCL), and incomplete reconstruction of adjacency matrix by SVD (\ie LightGCL) could disrupt the collaborative information within contrastive views~\cite{DBLP:conf/www/LinTHZ22,DBLP:conf/sigir/YangWW0HZZW23,DBLP:conf/sigir/LiXRY0023}, and alignment based on these contrastive views may mislead model learning in graph CL.
% In contrast, our approach aims to introduce discrete codes rich in collaborative information to enhance the neighborhood structure and semantic relevance of contrastive views.

\begin{figure}[]
\centering
\includegraphics[width=0.98\linewidth]{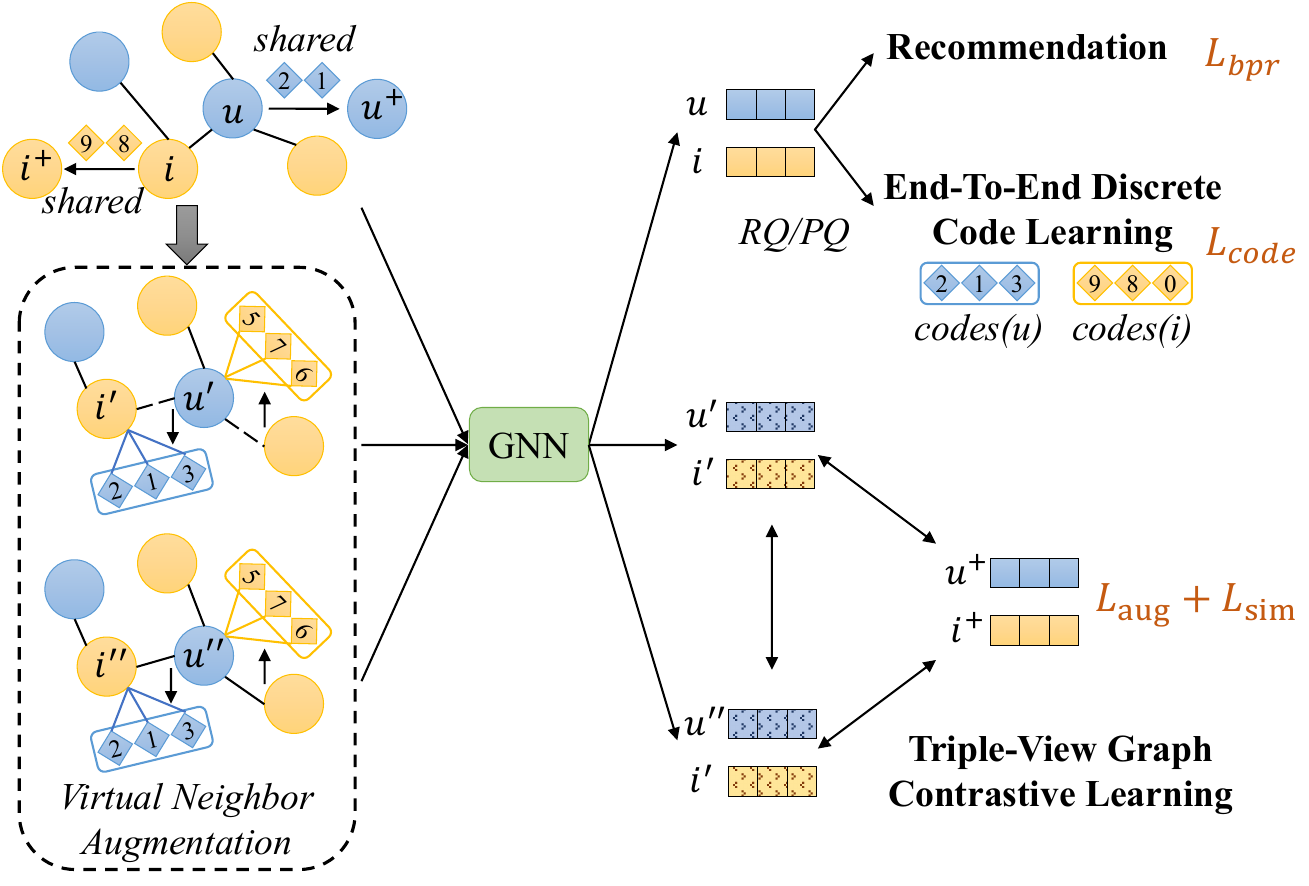}
\caption{The overall framework of our CoGCL, which enhances graph CL by constructing contrastive views that imply stronger collaborative information via discrete codes.}
\label{fig:model}
\end{figure}

\section{Methodology}
\label{sec:method}

In this section, we present our proposed CoGCL, a novel framework to enhance graph CL by constructing contrastive views that imply stronger collaborative information via discrete codes. The overall framework of our proposed approach is illustrated in Figure~\ref{fig:model}.

\subsection{Approach Overview}
\label{sec:overview}

As mentioned in Sections~\ref{sec:intro} and~\ref{sec:preliminary},  our basic idea is to enhance contrastive view generation and improve graph CL by introducing discrete codes associated with rich collaborative information.
To this end, we make efforts in the following aspects:

$\bullet$ \textbf{End-To-End Discrete Code Learning}~(Section~\ref{sec:code_learn}): In order to elegantly learn discrete codes associated with rich collaborative information to represent users and items, we present an end-to-end multi-level vector quantizer, which quantizes user and item representations encoded by GNN into discrete codes.

$\bullet$ \textbf{Reliable and Informative Contrastive View Generation}~(Section~\ref{sec:aug_sim}): Given the learned discrete codes, we use them for reliable and informative contrastive views by proposing virtual neighbor augmentation and semantic relevance sampling, respectively.

$\bullet$ \textbf{Triple-View Graph Contrastive Learning}~(Section~\ref{sec:cl}): Based on the generated contrastive views, we finally introduce triple-view graph contrastive learning to achieve alignment across multiple contrastive views, so as to integrate the stronger collaborative information contained in these views into model learning.

% Next, we elaborate on our methodology.

\subsection{End-To-End Discrete Code Learning}
\label{sec:code_learn}
% In this section, we introduce the approach for end-to-end discrete code learning to capture semantic connections between different users/items.
% To this end, we begin by encoding user and item representations that contain collaborative semantics and then leverage a multi-level vector quantization method to obtain user and item discrete codes.

As introduced before, we aim to learn discrete codes rich in collaborative information for users and items to enhance contrastive view generation. 
This involves (a) encoding user and item representations via GNN~(Section~\ref{sec:gnn}), and (b) learning end-to-end multi-level vector quantizer to map the encoded representations into discrete codes~(Section~\ref{sec:vq_code}).

\subsubsection{Representation Encoding via GNN}
\label{sec:gnn}

In line with previous works \cite{DBLP:conf/sigir/WuWF0CLX21,DBLP:conf/www/LinTHZ22,DBLP:conf/sigir/YuY00CN22}, we adopt LightGCN~\cite{DBLP:conf/sigir/0001DWLZ020} as the GNN encoder in our framework to propagate neighbor information across interaction graph due to its simplicity and effectiveness.
Notably, unlike previous implementations, we incorporate dropout on the input representation of each layer (instead of edge dropout on the graph structure) to mitigate overfitting. 
The process can be written as:
\begin{align}
    \mathbf{Z}^l &= \operatorname{GNN}(\rho(\mathbf{Z}^{l-1}), \mathcal{G}),
    \label{eq:gnn_enc}
\end{align}
where $\rho(\cdot)$ denotes the dropout operation. As for the readout function, we follow SimGCL~\cite{DBLP:conf/sigir/YuY00CN22} to skip $\mathbf{Z}^{0}$, which shows slight performance improvement in graph CL-based CF.
Subsequently, the user and item representations are denoted as $z_u$ and $z_i$, respectively, which will be applied for joint learning of the recommendation task and multi-level code.

\subsubsection{End-To-End Multi-Level Code Learning}
\label{sec:vq_code}

Given user and item representations, common approaches for learning discrete codes include hierarchical clustering~\cite{DBLP:journals/widm/MurtaghC12,DBLP:journals/corr/abs-2309-13375}, semantic hashing~\cite{DBLP:conf/stoc/Charikar02}, and vector quantization~\cite{gray1984vector,DBLP:conf/nips/RajputMSKVHHT0S23}. 
Our CoGCL adopts the multi-level vector quantization~(VQ) method in an end-to-end manner, such as residual quantization~(RQ)~\cite{DBLP:journals/sensors/ChenGW10} and product quantization~(PQ)~\cite{DBLP:journals/pami/JegouDS11}.
Next, we take discrete code learning for users as an example, and item codes can be obtained analogously.
At each level $h$, there exists a codebook $\mathcal{C}^h = \{\mathbf{e}_k^h\}_{k=1}^K$, where each vector $\mathbf{e}_k^h$ is a learnable cluster center. And the total number of code levels is $H$. 
Then the quantization process can be expressed as:
\begin{align}
    c_u^h = \underset {k} { \operatorname {arg\,max} } \  P(k|\mathbf{z}_u^h), \ \ \ \ 
    P(k|\mathbf{z}_u^h) = \frac{e^{s(\mathbf{z}_u^h,\mathbf{e}_k^h) / \tau} }{\sum_{j=1}^K e^{s(\mathbf{z}_u^h,\mathbf{e}_j^h) / \tau} },
    \label{eq:vq}
\end{align}
where $c_u^h$ is the $h$-th code for the user, $\mathbf{z}_u^h$ denotes user representation at the $h$-th level.
% Given the user representation $\mathbf{z}_u$ readout from GNN in Eq.~\eqref{eq:gnn}, 
RQ calculates residuals as representations for each level, denoted by $\mathbf{z}_u^{h+1} = \mathbf{z}_u^h - \mathbf{e}_{c_h}^h$, and $\mathbf{z}_u^1 = \mathbf{z}_u$.
PQ splits $\mathbf{z}_u$ into $H$ sub-vectors $\mathbf{z}_u = \left[\mathbf{z}_u^1;\dots; \mathbf{z}_u^H  \right]$, each of dimension $d/H$.
Here we do not adopt the Euclidean distance commonly used in prior VQ works~\cite{gray1984vector,vasuki2006review,DBLP:conf/nips/RajputMSKVHHT0S23,DBLP:journals/corr/abs-2311-09049} but cosine similarity, which is to synchronize with the similarity measure in CL~(Eq.~\eqref{eq:infonce}). 
% Furthermore, recent studies~\cite{DBLP:conf/iclr/YuLKZPQKXBW22,DBLP:conf/nips/KumarSLKK23} also demonstrate the merits of cosine similarity in terms of stability and quality.
% For similar purposes, 

Our optimization objective is to maximize the likelihood of assigning representations to their corresponding centers via Cross-Entropy (CE) loss.
% instead of Mean Squared Error (MSE) loss. 
Formally, the training loss for user discrete code learning is:
\begin{align}
    \mathcal{L}_{code}^U = -\frac{1}{H} \sum_{h=1}^H \log P(c_u^h|\mathbf{z}_u^h),
    \label{eq:code}
\end{align}
where $\mathcal{L}_{code}^U$ denotes the discrete code loss on the user side, and the loss for items is defined similarly, denoted by $\mathcal{L}_{code}^I$.
The total discrete code loss is formulated as $\mathcal{L}_{code} = \mathcal{L}_{code}^U + \mathcal{L}_{code}^I $.

\subsection{Reliable and Informative Contrastive View Generation}
\label{sec:aug_sim}
Compared to previous methods~\cite{DBLP:conf/sigir/WuWF0CLX21,DBLP:conf/iclr/Cai0XR23,DBLP:conf/sigir/YuY00CN22} involving information disruption, our motivation to strengthen collaborative information requires us to develop a reliable and informative approach for contrastive view generation via the learned discrete codes. 
% that leverages  rich in collaborative information . 
Below, we introduce virtual neighbor augmentation~(Section~\ref{sec:aug}) and semantic relevance sampling~(Section~\ref{sec:sim_samp}) to enhance the neighborhood structure and semantic relevance of contrastive views, respectively.
% Based on the learned discrete codes associated with rich collaborative information, we further use them to enhance contrastive view generation from 
% for virtual neighbor augmentation and semantic similarity sampling. 
% The former is to construct augmented nodes with common virtual neighbors that exhibit structural relevance.
% The latter is to obtain distinct users/items with similar collaborative semantics that exhibit semantic relevance.

\subsubsection{Virtual Neighbor Augmentation via Discrete Codes}
\label{sec:aug}

In order to generate reliable contrastive views with enhanced neighborhood structure, we use discrete codes for virtual neighbor augmentation in the graph.
For instance, considering user $u$, we select nodes from the user's neighbors $\mathcal{N}_u$ with a probability of $p$ to create augmented data, denoted as $\mathcal{N}_u^{\text{aug}}$.
Then we design two operators on graph structure to augment the node neighbors, \ie ``\emph{replace}'' and ``\emph{add}''.
The former replaces the neighbor items with their corresponding codes, without retaining the original edges, while the latter directly adds the codes as virtual neighbors. 
All augmentation operations strictly rely on observed interactions to ensure reliability.
Formally, the augmented edge of $u$ can be expressed as:
\begin{align}
    \mathcal{E}_u^{c} &= \left\{(u,c_i^{h})| i \in \mathcal{N}_u^{\text{aug}}, h \in \{1,\dots,H\}\right\}, \\
    \mathcal{E}_u^{r} &= \left\{(u,i)| i \in (\mathcal{N}_u \setminus \mathcal{N}_u^{\text{aug}}) \right\} \cup \mathcal{E}_u^{c}, \\
    \mathcal{E}_u^{a} &= \left\{(u,i)| i \in \mathcal{N}_u \right\} \cup \mathcal{E}_u^{c},
\end{align}
where $\mathcal{E}_u^{c}$ denotes the edges between user $u$ and discrete codes, $\mathcal{E}_u^{r}$ is all interaction edges of the user with ``\emph{replace}'' augmentation, and $\mathcal{E}_u^{a}$ is edges with ``\emph{add}'' augmentation.
In this case, discrete codes can be regarded as virtual neighbors of the user. 
The operations described above, which entail either replacing the original neighbor with several virtual neighbors or adding extra virtual neighbors, can bring richer neighbor information and effectively alleviate the sparsity of the graph.
The graph augmentation for items can be symmetrically performed.
To acquire a pair of augmented nodes for CL, we perform two rounds of virtual neighbor augmentation. The augmented graphs are depicted as follows:
\begin{align}
    \mathcal{G}^{1}=(\widetilde{\mathcal{V}}, \mathcal{E}^{o_1}), \ \ \ \
    \mathcal{G}^{2}=(\widetilde{\mathcal{V}}, \mathcal{E}^{o_2}), \ \ \ \
    o_1, o_2 \in \{r,a\}
\end{align}
where the node set $\widetilde{\mathcal{V}}=\{\mathcal{U} \cup \mathcal{C}^U  \cup \mathcal{I} \cup \mathcal{C}^I\}$ comprises all users, items and their corresponding discrete codes. 
Two stochastic operators $o_1$ and $o_2$ are selected from ``\emph{replace}''~(\ie $r$) and ``\emph{add}''~(\ie $a$).
$\mathcal{E}^{o_1}$ and $\mathcal{E}^{o_2}$ denote the edge sets resulting from the aforementioned virtual neighbor augmentation for all users and items.
% In the two augmented graphs obtained, the same node is deemed structurally relevant due to highly overlapping common neighbors (many code virtual neighbors).
The augmented nodes in the two graphs possess abundant (extensive virtual neighbors) and homogeneous (substantial common neighbors) neighbor structural information. 
Alignment between the two augmented nodes is helpful to introduce more neighbor structure information into the model.
Following SGL~\cite{DBLP:conf/sigir/WuWF0CLX21}, we update the discrete codes and augmented graphs once per epoch during training.

\subsubsection{Semantic Relevance Sampling via Discrete Codes}
\label{sec:sim_samp}

In our framework, we not only consider different augmented views of the same node as positive samples, but also regard distinct users/items with similar semantics as mutually positive, which leads to a more informative contrastive view.
This emphasizes the alignment of similar instances, rather than indiscriminately distancing different ones~\cite{DBLP:conf/sigir/XiaHXZYH22,DBLP:conf/sigir/YuY00CN22}.
Notably, different from NCL~\cite{DBLP:conf/www/LinTHZ22}, which learns cluster centers based on the EM algorithm as anchors, we measure semantic relevance in a more fine-grained manner based on discrete codes.
Specifically, we assess the semantic relevance of users in two ways:  
(a) \emph{Shared codes}: 
The discrete codes we learned are correlated with the collaborative semantics of user representations. 
Sharing codes between two users indicates fine-grained semantic relevance. Thus, we identify users who share at least $H$-1 codes as positive. 
(b) \emph{Shared target}: 
When two users share a common interacted target, that is, they possess the same prediction label in the dataset, we also consider them to be relevant. 
This supervised positive sampling method has shown its effectiveness in various scenarios, including sentence embedding~\cite{DBLP:conf/emnlp/GaoYC21} and sequential recommendation~\cite{DBLP:conf/wsdm/QiuHYW22}.
Given the positive set combined by the instances from the above two groups, we pair a sampled relevant instance with each user for CL.
Furthermore, semantically relevant positives of items can also be obtained in a symmetrical way.
By performing CL within the sampled instances above, we aim to enhance the clustering among similar users/items and improve semantic learning.

\subsection{Triple-View Graph Contrastive Learning}
\label{sec:cl}

After the above contrastive view generation methods, we can obtain three contrastive views with stronger collaborative information for each node through virtual neighbor augmentation and semantic relevance sampling: two augmented nodes with more abundant neighborhood structure and a semantically relevant user/item.
% To unleash the potential of graph CL, we propose performing cross-alignment between these triple views to integrate structural and semantic information effectively.
In this part, we first introduce how to encode multi-view node representations, and then present our triple-view graph contrastive learning approach to integrate structural and semantic information effectively.

\subsubsection{Multi-View Representation Encoding}
\label{sec:aug_gnn}

For the two augmented graphs, we introduce additional learnable embeddings of user and item discrete codes to serve as supplemental inputs, denoted as $\mathbf{Z}^c \in \mathbb{R}^{(|\mathcal{C}^U|+|\mathcal{C}^I|) \times d}$.
The input embedding matrix for augmented graphs is formed by concatenating ID embeddings with code embeddings, denoted as $\widetilde{\mathbf{Z}}^0 = [\mathbf{Z}^0;\mathbf{Z}^c]$.
Then we obtain representations of different views based on the same GNN encoder in Section~\ref{sec:gnn}:
\begin{align}
    \mathbf{Z}_1^l = \operatorname{GNN}(\rho(\mathbf{Z}_1^{l-1}), \mathcal{G}^1),\ \ \ \  \mathbf{Z}_2^l = \operatorname{GNN}(\rho(\mathbf{Z}_2^{l-1}), \mathcal{G}^2),
\end{align}
where the initial representations are set as $\mathbf{Z}_1^0 = \mathbf{Z}_2^0 = \widetilde{\mathbf{Z}}^0$.
After applying the readout function, we denote the representations of these two views as $\mathbf{Z}^{\prime}$, and $\mathbf{Z}^{\prime\prime}$, respectively.
As for the semantically relevant user/item, we directly adopt the node representation obtained based on the initial interaction graph in Section~\ref{sec:gnn} due to no structural augmentation.
Moreover, the representation dropout we introduced can also be regarded as a minor data augmentation.
The distinct dropout masks applied during the two forward propagations result in different features~\cite{DBLP:conf/emnlp/GaoYC21,DBLP:conf/cikm/YaoYCYCMHCTKE21,DBLP:conf/wsdm/QiuHYW22,DBLP:journals/tors/ZhouSLZM23}.

\subsubsection{Alignment Between Neighbor Augmented Views}
\label{sec:augcl}
As detailed in Section~\ref{sec:aug}, the two augmented nodes resulting from two rounds of virtual neighbor augmentation possess abundant neighbor structures. 
Therefore, we aim to incorporate more structural information and improve model efficacy by aligning these neighbor augmented views.
Formally, the alignment objective on the user side is as follows:
\begin{align}
    \mathcal{L}_{aug}^U = - \left(\log \frac{e^{s(\mathbf{z}_u^\prime,\mathbf{z}_u^{\prime \prime}) / \tau} }{\sum_{ \tilde{u} \in \mathcal{B} } e^{s(\mathbf{z}_u^\prime,\mathbf{z}_{\tilde{u}}^{\prime \prime}) / \tau} } 
    + \log \frac{e^{s(\mathbf{z}_u^{\prime \prime},\mathbf{z}_u^\prime) / \tau} }{\sum_{ \tilde{u} \in \mathcal{B} } e^{s(\mathbf{z}_u^{\prime \prime},\mathbf{z}_{\tilde{u}}^\prime) / \tau} } \right),
    \label{eq:augcl}
\end{align}
where $u$ and $\tilde{u}$ are users in batch data $\mathcal{B}$. 
$\mathbf{z}_u^\prime$ and $\mathbf{z}_u^{\prime \prime}$ denote two different user representations after virtual neighbor augmentations. 
The loss consists of two terms, representing the bidirectional alignment of the two views.
Analogously, we calculate the CL loss for the item side as $\mathcal{L}_{aug}^I$. 
The total alignment loss between nodes with augmented views is the sum of them, denoted as $\mathcal{L}_{aug} = \mathcal{L}_{aug}^U + \mathcal{L}_{aug}^I$.

\subsubsection{Alignment Between Semantically Relevant Users/Items}
\label{sec:simcl}
Following the semantics relevance sampling method in Section~\ref{sec:sim_samp}, we randomly select a positive example with similar collaborative semantics for each user $u$, denoted as $u^+$. 
Then we align these relevant users to incorporate more collaborative semantic information into the model.
The alignment loss can be written as:
\begin{align}
    \mathcal{L}_{sim}^U = - \left(\log \frac{e^{s(\mathbf{z}_u^\prime,\mathbf{z}_{u^+}) / \tau} }{\sum_{ \tilde{u} \in \widetilde{\mathcal{B}} } e^{s(\mathbf{z}_u^\prime,\mathbf{z}_{\tilde{u}}) / \tau} } 
    + \log \frac{e^{s(\mathbf{z}_u^{\prime\prime},\mathbf{z}_{u^+}) / \tau} }{\sum_{ \tilde{u} \in \widetilde{\mathcal{B}} } e^{s(\mathbf{z}_u^{\prime\prime},\mathbf{z}_{\tilde{u}}) / \tau} } \right),
    \label{eq:simcl}
\end{align}
where $(u,u^+)$ is a positive user pair, and $\widetilde{\mathcal{B}}$ is the sampled data in a batch.
The two components of the equation correspond to the alignment between two augmented views and the similar user, respectively.
Furthermore, combining the symmetric alignment loss on the item side, the total alignment loss between similar users/items is $\mathcal{L}_{sim}=\mathcal{L}_{sim}^U+\mathcal{L}_{sim}^I$.

\subsubsection{Overall Optimization}

In the end, by combining the recommendation loss~(\ie BPR loss), discrete code learning objective~(Eq.~\eqref{eq:code}) and all contrastive learning loss~(Eq.~\eqref{eq:augcl} and Eq.~\eqref{eq:simcl}), our CoGCL is jointly optimized by minimizing the following overall loss:
\begin{align}
    \mathcal{L} = \mathcal{L}_{bpr} + \lambda \mathcal{L}_{code} + \mu \mathcal{L}_{aug} + \eta \mathcal{L}_{sim},
    \label{eq:loss}
\end{align}
where $\lambda$, $\mu$ and $\eta$ are hyperparameters for the trade-off between various objectives.

\subsection{Discussion}
\label{sec:discussion}

In this section, we make a brief comparison with existing graph CL-based CF methods to highlight the novelty and contributions of CoGCL. 
According to how to construct contrast views, existing methods can be divided into two categories: structure augmentation and representation augmentation. 

\paratitle{Structural augmentation methods} typically generate contrastive views by perturbing the graph structure like stochastic node/edge dropout~\cite{DBLP:conf/sigir/WuWF0CLX21}. 
Several recent efforts attempt to use well-founded methods for structural perturbations, such as SVD-based adjacency matrix reconstruction~\cite{DBLP:conf/iclr/Cai0XR23} and graph rationale discovery based on masked autoencoding~\cite{DBLP:conf/sigir/LiXRY0023}. 
However, perturbations on sparse graphs can not construct more informative contrastive views.
As a comparison, our approach is both \emph{reliable} and \emph{informative}, leveraging discrete codes as virtual neighbors to reliably enhance node neighborhood structure and alleviate data sparsity. 
The alignment between two augmented nodes with abundant neighbors is beneficial for the integration of further collaborative information.

\paratitle{Representation augmentation methods} involve modeling additional node representations as contrastive views, such as learning hypergraph representations~\cite{DBLP:conf/sigir/XiaHXZYH22} and adding random noise~\cite{DBLP:conf/sigir/YuY00CN22}. 
However, limited by the low-rank hypergraph matrix and the noise perturbation, the generated contrastive views also suffer from the semantic disruption issue.
Besides, these methods typically indiscriminately distinguish representations of different instances.
In contrast, we consider users/items with shared codes or interaction targets as semantically relevant.
By aligning users/items with similar collaborative semantics, we can further unleash the potential of CL and enhance the semantic learning of the model.

\begin{table}[]
\centering
\caption{Statistics of the preprocessed datasets.}
\label{tab:data_statistics}
\begin{tabular}{lrrrr}
\toprule
 \textbf{Datasets} & \textbf{\#Users} & \textbf{\#Items}  & \textbf{\#Interactions} & \textbf{Sparsity} \\
\midrule
Instrument & 48,453& 21,413  & 427,674  & 99.959\% \\
Office & 181,878 & 67,409 & 1,477,820   & 99.988\%  \\
Gowalla & 29,858 & 40,988 & 1,027,464 & 99.916\%  \\ 
iFashion  & 300,000 & 81,614 & 1,607,813 & 99.993\%  \\ 
\bottomrule
\end{tabular}
\end{table}

\begin{table*}[]
\centering
\caption{Performance comparison of different methods on the four datasets. The best and second-best performances are indicated in bold and underlined font, respectively. }
\label{tab:res}
\resizebox{\textwidth}{!}{
\begin{tabular}{l|l|cccc@{\hspace{1.5mm}}c@{\hspace{1.5mm}}c|cccc@{\hspace{2mm}}c@{\hspace{1.5mm}}c@{\hspace{1.5mm}}c|cc}
\hline
\multicolumn{1}{l|}{Dataset}  & Metric    & BPR    & GCMC   & NGCF   & DGCF   & LightGCN & SimpleX & SLRec  & SGL    & NCL    & HCCF   & GFormer & SimGCL & LightGCL & CoGCL  & Improv. \\
\hline
\multirow{6}{*}{Instrument} & Recall@5  & 0.0293 & 0.0334 & 0.0391 & 0.0401 & 0.0435   & 0.0386  & 0.0381 & 0.0449       & 0.0449 & 0.0456 & \underline{0.0471} & 0.0470       & 0.0468       & \textbf{0.0515} & 9.34\%    \\
                             & NDCG@5    & 0.0194 & 0.0218 & 0.0258 & 0.0269 & 0.0288   & 0.0244  & 0.0256 & 0.0302       & 0.0302 & 0.0303 & 0.0314       & \underline{0.0316} & 0.0310       & \textbf{0.0345} & 9.18\%   \\
                             & Recall@10 & 0.0469 & 0.0532 & 0.0617 & 0.0628 & 0.0660   & 0.0631  & 0.0574 & 0.0692       & 0.0685 & 0.0703 & 0.0715       & \underline{0.0717} & 0.0715       & \textbf{0.0788} & 9.90\%    \\
                             & NDCG@10   & 0.0250 & 0.0282 & 0.0331 & 0.0342 & 0.0361   & 0.0324  & 0.0319 & 0.0380       & 0.0377 & 0.0383 & 0.0393       & \underline{0.0395} & 0.0391       & \textbf{0.0435} & 10.13\%   \\
                             & Recall@20 & 0.0705 & 0.0824 & 0.0929 & 0.0930 & 0.0979   & 0.0984  & 0.0820 & 0.1026       & 0.1011 & 0.1028 & 0.1041       & \underline{0.1057} & 0.1042       & \textbf{0.1152} & 8.99\%    \\
                             & NDCG@20   & 0.0310 & 0.0357 & 0.0411 & 0.0419 & 0.0442   & 0.0413  & 0.0381 & 0.0466       & 0.0459 & 0.0466 & 0.0478       & \underline{0.0482} & 0.0474       & \textbf{0.0526} & 9.13\%    \\
                             \hline
\multirow{6}{*}{Office}      & Recall@5  & 0.0204 & 0.0168 & 0.0178 & 0.0258 & 0.0277   & 0.0291  & 0.0294 & 0.0349       & 0.0293 & 0.0340 & \underline{0.0353} & 0.0349       & 0.0338       & \textbf{0.0411} & 16.43\%   \\
                             & NDCG@5    & 0.0144 & 0.0109 & 0.0116 & 0.0177 & 0.0186   & 0.0199  & 0.0209 & 0.0242       & 0.0201 & 0.0230 & \underline{0.0245} & 0.0240       & 0.0232       & \textbf{0.0287} & 17.14\%   \\
                             & Recall@10 & 0.0285 & 0.0270 & 0.0279 & 0.0380 & 0.0417   & 0.0422  & 0.0402 & 0.0493       & 0.0434 & 0.0489 & 0.0492       & \underline{0.0494} & 0.0490       & \textbf{0.0582}  & 17.81\%   \\
                             & NDCG@10   & 0.0170 & 0.0141 & 0.0149 & 0.0217 & 0.0231   & 0.0241  & 0.0244 & 0.0289       & 0.0243 & 0.0282 & \underline{0.0292} & 0.0289       & 0.0280       & \textbf{0.0343} & 17.47\%   \\
                             & Recall@20 & 0.0390 & 0.0410 & 0.0438 & 0.0544 & 0.0605   & 0.0602  & 0.0534 & 0.0681       & 0.0629 & 0.0677 & 0.0672       & 0.0689       & \underline{0.0698} & \textbf{0.0785} & 12.46\%   \\
                             & NDCG@20   & 0.0197 & 0.0178 & 0.0189 & 0.0258 & 0.0279   & 0.0287  & 0.0277 & 0.0336       & 0.0292 & 0.0331 & \underline{0.0338} & 0.0337       & 0.0332       & \textbf{0.0393} & 14.18\%   \\
                             \hline
\multirow{6}{*}{Gowalla}     & Recall@5  & 0.0781 & 0.0714 & 0.0783 & 0.0895 & 0.0946   & 0.0782  & 0.0689 & \underline{0.1047} & 0.1040 & 0.0836 & 0.1042       & \underline{0.1047} & 0.0947       & \textbf{0.1092} & 4.30\%    \\
                             & NDCG@5    & 0.0707 & 0.0633 & 0.0695 & 0.0801 & 0.0854   & 0.0712  & 0.0613 & 0.0955       & 0.0933 & 0.0749 & 0.0935       & \underline{0.0959} & 0.0860       & \textbf{0.0995} & 3.75\%    \\
                             & Recall@10 & 0.1162 & 0.1089 & 0.1150 & 0.1326 & 0.1383   & 0.1187  & 0.1045 & 0.1520       & 0.1508 & 0.1221 & 0.1515       & \underline{0.1525} & 0.1377       & \textbf{0.1592} & 4.39\%    \\
                             & NDCG@10   & 0.0821 & 0.0749 & 0.0808 & 0.0932 & 0.0985   & 0.0834  & 0.0722 & 0.1092       & 0.1078 & 0.0866 & 0.1085       & \underline{0.1100} & 0.0988       & \textbf{0.1145} & 4.09\%    \\
                             & Recall@20 & 0.1695 & 0.1626 & 0.1666 & 0.1914 & 0.2002   & 0.1756  & 0.1552 & 0.2160       & 0.2130 & 0.1794 & 0.2166       & \underline{0.2181} & 0.1978       & \textbf{0.2253} & 3.30\%    \\
                             & NDCG@20   & 0.0973 & 0.0903 & 0.0956 & 0.1100 & 0.1161   & 0.0996  & 0.0868 & 0.1274       & 0.1254 & 0.1029 & 0.1271       & \underline{0.1286} & 0.1159       & \textbf{0.1333} & 3.65\%    \\
                             \hline
\multirow{6}{*}{iFashion}    & Recall@5  & 0.0195 & 0.0240 & 0.0234 & 0.0297 & 0.0309   & 0.0345  & 0.0237 & 0.0377       & 0.0330 & 0.0419 & 0.0354       & 0.0401       & \underline{0.0423} & \textbf{0.0463} & 9.46\%    \\
                             & NDCG@5    & 0.0128 & 0.0156 & 0.0151 & 0.0197 & 0.0205   & 0.0231  & 0.0157 & 0.0252       & 0.0219 & 0.0280 & 0.0235       & 0.0267       & \underline{0.0284} & \textbf{0.0310}  & 9.15\%    \\
                             & Recall@10 & 0.0307 & 0.0393 & 0.0384 & 0.0459 & 0.0481   & 0.0525  & 0.0361 & 0.0574       & 0.0501 & 0.0636 & 0.0540       & 0.0608       & \underline{0.0641} & \textbf{0.0696} & 8.58\%    \\
                             & NDCG@10   & 0.0164 & 0.0206 & 0.0199 & 0.0249 & 0.0260   & 0.0289  & 0.0198 & 0.0315       & 0.0274 & 0.0350 & 0.0294       & 0.0334       & \underline{0.0354} & \textbf{0.0386} & 9.04\%    \\
                             & Recall@20 & 0.0470 & 0.0623 & 0.0608 & 0.0685 & 0.0716   & 0.0770  & 0.0535 & 0.0846       & 0.0742 & 0.0929 & 0.0790       & 0.0897       & \underline{0.0932} & \textbf{0.1010}  & 8.37\%    \\
                             & NDCG@20   & 0.0206 & 0.0264 & 0.0256 & 0.0307 & 0.0320   & 0.0351  & 0.0242 & 0.0384       & 0.0335 & 0.0425 & 0.0358       & 0.0407       & \underline{0.0428} & \textbf{0.0465} & 8.64\%    \\
                             \hline
\end{tabular}
}
\end{table*}

\section{Experiment}
% In this section, we begin with the detailed experiment setup and then present overall performance and in-depth analysis of our proposed approach.

\subsection{Experiment Setup}

\subsubsection{Dataset}
We evaluate our proposed approach on four public datasets:
Instrument and Office subsets from the most recent Amazon2023 benchmark~\cite{DBLP:journals/corr/abs-2403-03952}, Gowalla~\cite{DBLP:conf/kdd/ChoML11}, Alibaba-iFashion~\cite{DBLP:conf/kdd/ChenHXGGSLPZZ19}.
For Instrument and Office datasets, we filter out low-activity users and items with less than five interactions.
For Gowalla dataset, we use 10-core filtering to ensure the data quality following prior works~\cite{DBLP:conf/sigir/Wang0WFC19,DBLP:conf/sigir/0001DWLZ020}.
As for the sparser iFashion dataset, we employ the data processed by \cite{DBLP:conf/sigir/WuWF0CLX21}, which randomly samples 300k users and their interactions.
Our processed datasets vary in terms of domain, scale, and sparsity. 
Their statistics are summarized in Table~\ref{tab:data_statistics}.
For each dataset, we split the observed interactions into training, validation, and testing sets with a ratio of 8:1:1.

\subsubsection{Baseline Models}
We adopt the following competitive baselines for comparison with our CoGCL,  
which includes \textbf{traditional CF models}: 
(1) {\textbf{BPR}}~\cite{DBLP:conf/uai/RendleFGS09},
(2) {\textbf{GCMC}}~\cite{DBLP:journals/corr/BergKW17},
(3) {\textbf{NGCF}}~\cite{DBLP:conf/sigir/Wang0WFC19},
(4) {\textbf{DGCF}}~\cite{DBLP:conf/sigir/WangJZ0XC20},
(5) {\textbf{LightGCN}}~\cite{DBLP:conf/sigir/0001DWLZ020},
(6) {\textbf{SimpleX}}~\cite{DBLP:conf/cikm/MaoZWDDXH21}, 
as well as various representative \textbf{CL-based models}:
(7) {\textbf{SLRec}}~\cite{DBLP:conf/cikm/YaoYCYCMHCTKE21},
(8) {\textbf{SGL}}~\cite{DBLP:conf/sigir/WuWF0CLX21},
(9) {\textbf{NCL}}~\cite{DBLP:conf/www/LinTHZ22},
(10) {\textbf{HCCF}}~\cite{DBLP:conf/sigir/XiaHXZYH22},
(11) {\textbf{GFormer}}~\cite{DBLP:conf/sigir/LiXRY0023},
(12) {\textbf{SimGCL}}~\cite{DBLP:conf/sigir/YuY00CN22},
(13) {\textbf{LightGCL}}~\cite{DBLP:conf/iclr/Cai0XR23}.
A more detailed introduction to the above baseline models is given in Appendix~\ref{apd:baseline}.

\subsubsection{Evaluation Settings}
To evaluate the performance of the above models, we adopt two widely used metrics in recommendation: Recall@$N$ and Normalized Discounted Cumulative Gain (NDCG)@$N$. 
In this paper, we set $N$ to 5, 10, and 20.
For the sake of rigorous comparison, we perform full ranking evaluation~\cite{DBLP:conf/cikm/ZhaoCWGW20,DBLP:journals/tois/ZhaoLFWW23} over the entire item set instead of sample-based evaluation.

\subsubsection{Implementation Details}
% We implement our CoGCL and all baselines with RecBole~\cite{DBLP:conf/cikm/ZhaoMHLCPLLWTMF21,DBLP:conf/cikm/ZhaoHPYZLZBTSCX22}, which is a user-friendly open-source library for recommender systems.
For all comparison models, we use Adam for optimization and set the embedding dimension to 64 uniformly.
The batch size is 4096, and the number of GNN layers in GNN-based methods is set to 3.
To ensure a fair comparison, we utilize grid search to obtain optimal performance according to the hyperparameter settings reported in the original papers of baseline methods.
For our approach, we employ RQ as the default discrete code learning method. The number of code levels $H=4$, and the temperature $\tau=0.2$. 
The codebook size $K$ is set to 256 for Instrument and Gowalla datasets, and 512 for Office and iFashion datasets due to their larger scale.
The hyperparameters $\lambda$ are tuned in \{5, 1, 0.5\}, while $\mu$ and $\eta$ are tuned in \{5, 1, 0.5, 0.2, 0.1, 0.05, 0.02, 0.01, 0.005, 0.001\}.
The probabilities of ``\emph{replace}'' and ``\emph{add}'' in virtual neighbor augmentation are tuned in \{0.01, 0.05, 0.1, 0.15, 0.2, 0.25, 0.3, 0.4, 0.5, 0.6\}.
For experiments on hyperparameter tuning, please refer to Appendix~\ref{apd:hyper_tune}.
% And more implementation details can be found in our code\footnote{\url{https://anonymous.4open.science/r/CoGCL-D9EC/}}.

\subsection{Overall Performance}
The overall results for performance comparison between CoGCL and other baseline models are shown in Table~\ref{tab:res}.
From the results, we find the following observations:

The CL-based methods~(\eg SGL, NCL, SimGCL, LightGCL) show consistent superiority over the traditional MF methods~(\eg BPR, SimpleX) and GNN-based methods~(\eg NGCF, LightGCN).
This performance improvement could be attributed to the self-supervised signals brought by contrastive learning, which helps to alleviate data sparsity and enhance representation learning.
Within CL-based methods, structure augmentation and representation augmentation exhibit distinct strengths in different scenarios.
Specifically, SimGCL, as a typical representation augmentation method, performs better than other baseline models on Instrument and Gowalla datasets, thanks to the improved uniformity achieved by incorporating random noise.
Conversely, the most competitive models for Office and iFashion datasets are GFormer and LightGCL, respectively, both of which are structure augmentation methods. 
In contrast, SGL tends to underperform, indicating that stochastic edge/node dropout possibly interferes with crucial structural information, leading to adverse impacts.

Finally, our proposed CoGCL consistently maintains the best performance in all cases, achieving significant improvements over baseline methods. 
Different from these baseline models, CoGCL unleashes the potential of CL by constructing contrastive views that imply stronger collaborative information.
Based on the learned discrete codes rich in collaborative information, we introduce virtual neighbor augmentation and semantic relevance sampling to enhance the neighborhood structure and semantic relevance of contrasting views, respectively.
Furthermore, triple-view graph contrastive learning across the obtained contrastive views brings supplemental collaborative insights to the model. 
As a result, CoGCL exhibits strong robustness and effectiveness on sparse datasets ~(\eg Office, iFashion).

\subsection{Ablation Study}
In this part, we first investigate the contribution of various contrastive view generation methods in the proposed approach, and then conduct an in-depth ablation analysis of \emph{alignment} and \emph{uniformity} of CL.

\begin{figure}[]
\centering
\includegraphics[width=0.99\linewidth]{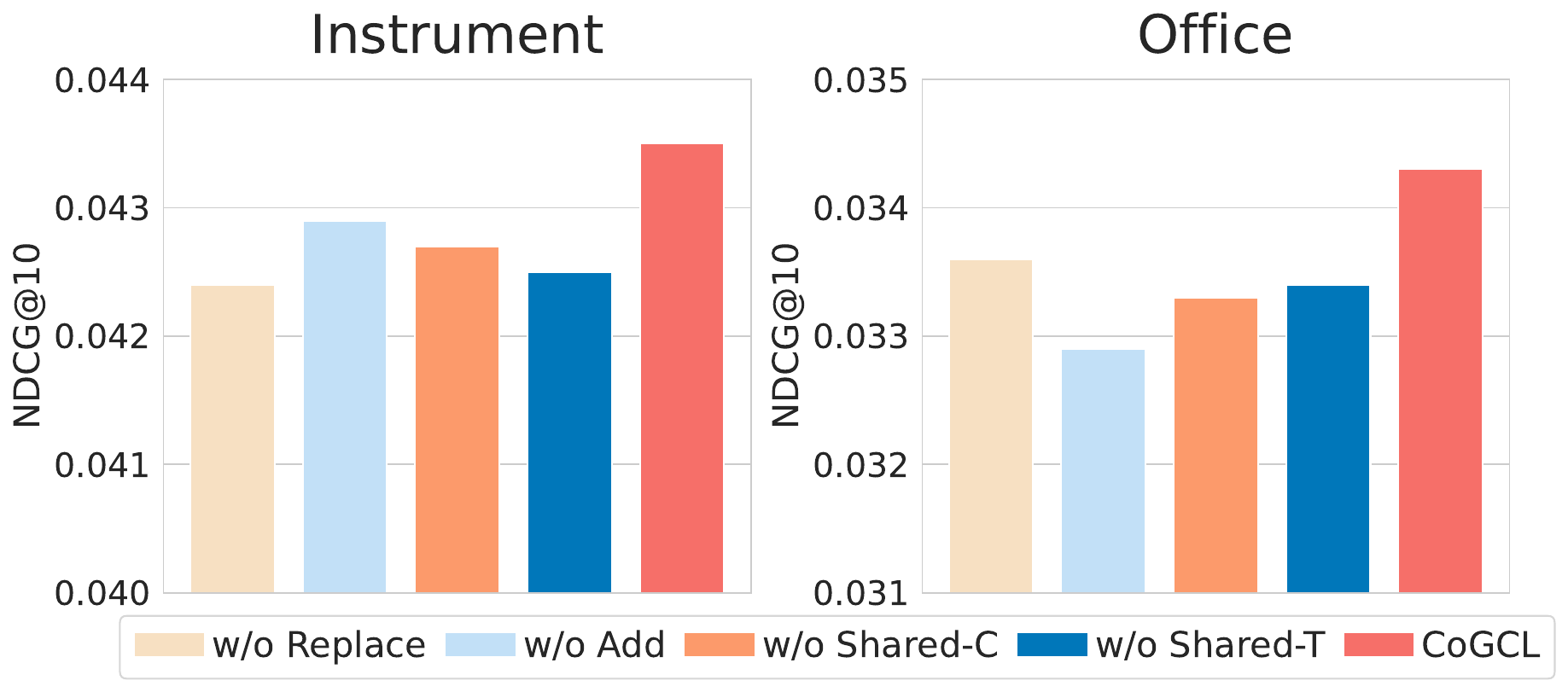}
\caption{Ablation study of data augmentation methods.}
\label{fig:aug_ablation}
\end{figure}

\subsubsection{Ablation Study of Data Augmentation}

In order to explore the contribution of data augmentation methods involved in CoGCL. 
we evaluate the performance of the following variants: 
(1) \underline{w/o Replace} removes the ``\emph{replace}'' operator in virtual neighbor augmentation. 
(2) \underline{w/o Add} removes the ``\emph{add}'' operator in virtual neighbor augmentation.
(3) \underline{w/o Shared-C} removes similar users/items shared codes in semantic relevance sampling.
(4) \underline{w/o Shared-T} removes similar users/items shared interaction target in semantic relevance sampling.
The results are shown in Figure~\ref{fig:aug_ablation}. 
We can observe that the exclusion of any data augmentation method would lead to a decrease in performance, which demonstrates that all data augmentation methods employed for contrastive view generation in CoGCL are useful for performance improvement.

\begin{table}[]
\centering
\caption{Performance analysis of alignment and uniformity in CoGCL.}
\label{tab:clcu}
\resizebox{\linewidth}{!}{
\renewcommand\arraystretch{0.9}
\begin{tabular}{l|cc|cc}
\hline
\multirow{2}{*}{Methods} & \multicolumn{2}{c|}{Instrument}    & \multicolumn{2}{c}{Office}           \\
\cline{2-5}
 & Recall@10       & NDCG@10       & Recall@10     & NDCG@10    \\
\hline
LightGCN         & 0.0660         & 0.0361        &  0.0417          & 0.0231           \\
CoGCL         & \textbf{0.0788}          & \textbf{0.0435}        & \textbf{0.0582}              & \textbf{0.0343}           \\
\ w/o A       & 0.0726          & 0.0401        & 0.0490              & 0.0280           \\
\ w/o U       & 0.0703          & 0.0384        & 0.0465              & 0.0267           \\
\ w/o AA      & 0.0741          & 0.0411        & 0.0536              & 0.0315          \\
\ w/o AU      & 0.0762          & 0.0421        & 0.0542              & 0.0306          \\
\ w/o SA      & 0.0767          & 0.0422        & 0.0554              & 0.0329          \\
\ w/o SU      & 0.0779          & 0.0429        & 0.0574              & 0.0336          \\

\hline
\end{tabular}
}
\end{table}

\subsubsection{Ablation Study of Triple-View Graph Contrastive Learning}

Apart from the above techniques, we further investigate how the \emph{alignment} and \emph{uniformity} of CL affect our approach.
We disable these two terms respectively in the CL losses~(\ie $\mathcal{L}_{aug}$ and $\mathcal{L}_{sim}$ in Section~\ref{sec:cl}) by applying the same gradient-stopping operations in empirical analysis~(Section~\ref{sec:emp_ana}). 
Specifically, we construct the following variants for detailed exploration:
(1) \underline{w/o A} and (2) \underline{w/o U} are consistent with Section~\ref{sec:emp_ana}, denoting disabling alignment and uniformity in CL respectively, including both $\mathcal{L}_{aug}$ and $\mathcal{L}_{sim}$.
(3) \underline{w/o AA} and (4) \underline{w/o AU} only involve disabling the above two terms of $\mathcal{L}_{aug}$ while keeping $\mathcal{L}_{sim}$ constant.
(5) \underline{w/o SA} and (6) \underline{w/o SU} are analogous variants for $\mathcal{L}_{sim}$ and do not change $\mathcal{L}_{sim}$.

As shown in Table~\ref{tab:clcu}, the absence of alignment~(\ie \underline{w/o A}) or uniformity~(\ie \underline{w/o U}) within both $\mathcal{L}_{aug}$ and $\mathcal{L}_{sim}$ leads to a notable performance degradation. 
This observation verifies that the joint effect of these two elements is crucial for the effectiveness of the proposed approach, rather than relying solely on uniformity.
Furthermore, individually disabling uniformity within $\mathcal{L}_{aug}$~(\ie \underline{w/o AU}) and $\mathcal{L}_{sim}$~(\ie \underline{w/o SU}) does not result in the significant adverse impact as conjectured.
% This phenomenon 
It could be attributed to the shared uniformity effect between the two CL losses in CoGCL, which may mutually reinforce each other.
In contrast, the individual deactivation of alignment within $\mathcal{L}_{aug}$~(\ie \underline{w/o AA}) and $\mathcal{L}_{sim}$~(\ie \underline{w/o SA}) incurs a pronounced decrease in performance.
This provides further evidence that our proposed alignment between the two types of positives brings enhanced collaborative information beyond uniformity.

\subsection{Further Analysis}

\begin{figure}[]
\centering
\includegraphics[width=0.99\linewidth]{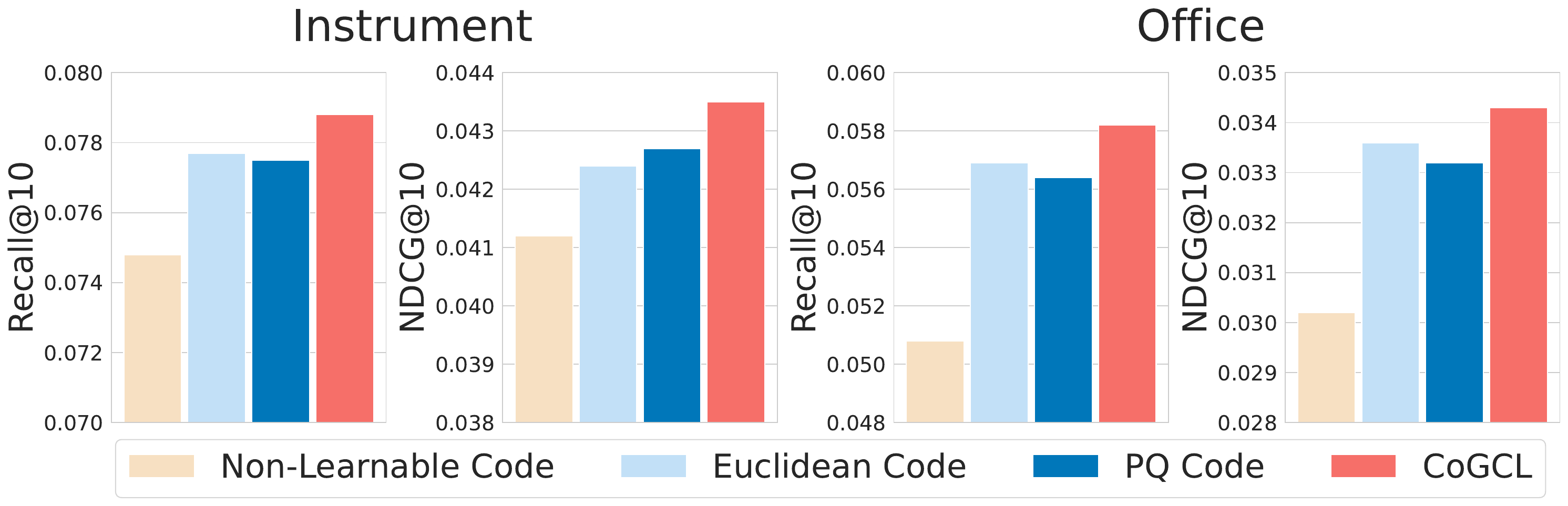}
\caption{Performance comparison of different discrete code learning methods.}
\label{fig:code_ablation}
\end{figure}

\subsubsection{Performance Comparison \wrt Different Discrete Code Learning Methods}
\label{apd:code_method}
To verify the advancedness of the proposed end-to-end discrete code learning method, we compare it with the following three variants:
(1) \underline{Non-Learnable Code} uses Faiss library~\cite{DBLP:journals/tbd/JohnsonDJ21} to generate discrete codes based on trained LightGCN embeddings. The generated codes are non-learnable and remain unchanged during model training.
(2) \underline{Euclidean Code} adopts Euclidean distance to measure the similarity between user/item representations and codebook vectors in Eq.~\eqref{eq:vq}, which is consistent with the original RQ method~\cite{DBLP:journals/sensors/ChenGW10}.
(3) \underline{PQ Code} employs PQ instead of RQ as a multi-level quantizer for discrete code learning.
We conduct experiments on Instrument and Office datasets, and the results are shown in Figure~\ref{fig:code_ablation}.
It can be seen that \underline{Non-Learnable Code} is less robust compared to the end-to-end learned discrete codes, which may stem from the inability to continuously improve the collaborative information within discrete codes while optimizing the model.
In comparison to \underline{Euclidean Code} and \underline{PQ Code}, our proposed approach shows superior performance. 
Unlike \underline{Euclidean Code}, our method utilizes cosine similarity to synchronize with the similarity measure in CL.
Compared with \underline{PQ Code}, the RQ we applied establishes conditional probability relationships among codes at each level instead of treating them as independent, which is conducive to the semantic modeling of various granularities.

\begin{figure}[]
\centering
\includegraphics[width=0.99\linewidth]{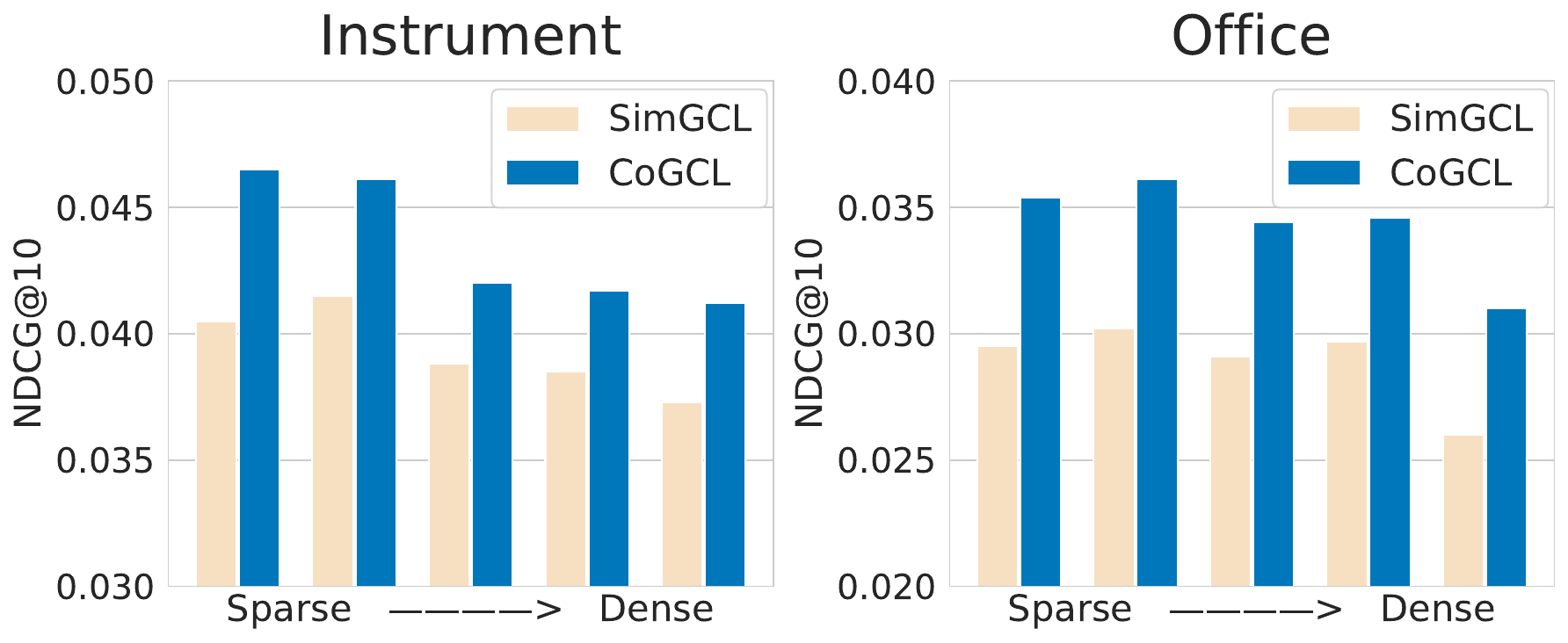}
\caption{Performance comparison on user groups with different sparsity levels.}
\label{fig:data_sparsity}
\end{figure}

\subsubsection{Performance Comparison \wrt Data Sparsity}
To verify the merit of our approach in alleviating data sparsity, we evaluate CoGCL on user groups with different sparsity levels.
Specifically, following prior works~\cite{DBLP:conf/www/LinTHZ22,DBLP:conf/iclr/Cai0XR23}, we divide users into five groups according to their number of interactions, while keeping the same number of users in each group constant.
Subsequently, we evaluate the performance of these five groups of users, and the results are shown in Figure~\ref{fig:data_sparsity}.
We can see that CoGCL consistently outperforms the baseline model across all sparsity levels. 
Furthermore, our model shows superior performance and significant improvement in the highly sparse user groups.
This phenomenon indicates that CoGCL can achieve high-quality recommendation in scenarios with sparse interactions, which benefits from the additional insights brought by CL between contrastive views with stronger collaborative information.

\section{Related Work}
\paratitle{GNN-Based Collaborative Filtering.}
Graph Neural Networks (GNNs) have become prominent in collaborative filtering~(CF) due to their effectiveness in modeling user-item relationships~\cite{DBLP:journals/csur/WuSZXC23,DBLP:journals/tors/GaoZLLQPQCJHL23}. 
The core approach involves organizing user-item interaction data into a bipartite graph and learning node representations from the graph structure.
Earlier efforts~\cite{DBLP:conf/www/BalujaSSJYKRA08,DBLP:conf/ijcai/GoriP07} extract the graph information using random walk strategies.
With the development of GNNs, the common studies has shifted towards designing effective message-passing mechanisms to propagate user/item embeddings over the graph~\cite{DBLP:journals/corr/BergKW17,DBLP:conf/kdd/YingHCEHL18,DBLP:conf/sigir/Wang0WFC19}.
Subsequently, LightGCN~\cite{DBLP:conf/sigir/0001DWLZ020} and LR-GCCF~\cite{DBLP:conf/aaai/ChenWHZW20} propose eliminating transformation and non-linear activation to simplify GNNs while improving performance.
Furthermore, recent studies are also devoted to enhancing GNNs with various advanced techniques, such as disentangled representation learning~\cite{DBLP:conf/sigir/WangJZ0XC20,DBLP:conf/cikm/WangTLSWZ20}, hypergraph learning~\cite{DBLP:conf/kdd/JiFJZT020,DBLP:conf/www/YuYLWH021} and contrastive learning~\cite{DBLP:conf/sigir/WuWF0CLX21,DBLP:conf/www/LinTHZ22,DBLP:conf/sigir/YuY00CN22,DBLP:conf/iclr/Cai0XR23}.

\paratitle{Contrastive Learning for Recommendation.}
Recently, contrastive learning~(CL) has demonstrated significant potential in various recommendation scenarios like sequential recommendation~\cite{DBLP:conf/cikm/ZhouWZZWZWW20,DBLP:conf/icde/XieSLWGZDC22,DBLP:conf/wsdm/QiuHYW22} and knowledge graph-enhanced recommendation~\cite{DBLP:conf/sigir/Zou0MWQ0C22,DBLP:conf/cikm/Zou0WM0FC22}.
% ,  click-through rate prediction~\cite{DBLP:journals/corr/abs-2109-13921,DBLP:journals/corr/abs-2306-02841,DBLP:journals/corr/abs-2311-15493}.
In the context of GNN-based CF, existing efforts can be categorized into two main approaches according to how the contrastive views are constructed.
The first approach is to perform data augmentation over graph structure~\cite{DBLP:conf/sigir/WuWF0CLX21,DBLP:conf/iclr/Cai0XR23,DBLP:conf/sigir/LiXRY0023,DBLP:conf/sigir/RenXZY023} 
For instance, SGL~\cite{DBLP:conf/sigir/WuWF0CLX21} randomly drops nodes/edges within the interaction graph to construct augmented graphs. 
The second approach is to model additional view representations of users and items for CL~\cite{DBLP:conf/sigir/LeeKJPY21,DBLP:conf/sigir/XiaHXZYH22,DBLP:conf/www/LinTHZ22,DBLP:conf/sigir/YuY00CN22,DBLP:journals/tkde/YuXCCHY24,DBLP:conf/kdd/TangDSCXYHJL24}. 
Particularly, SimGCL~\cite{DBLP:conf/sigir/YuY00CN22} generates contrastive views by adding random noise to node embeddings.
Despite their success, the collaborative information within contrastive views may be disrupted in these methods, and thus the potential of CL has not been fully exploited.
In this paper, we propose to unleash the potential of CL by constructing contrastive views with stronger collaborative information via discrete codes.

\paratitle{User/Item ID Discretization in Recommendation.}
ID discretization involves employing a tuple of discrete codes as identifier to represent a user/item instead of the vanilla single ID, achieved through methods like semantic hashing~\cite{DBLP:conf/stoc/IndykM98,DBLP:conf/stoc/Charikar02,DBLP:conf/acl/HenaoCSSWWC18}, vector quantization~\cite{gray1984vector,vasuki2006review}, etc. 
These methods allow similar users/items to share certain codes, which can offer valuable prior knowledge for subsequent recommendation models.
Initially, the focus was on developing memory- and time-efficient recommendation algorithms by sharing code embeddings~\cite{DBLP:conf/recsys/BalenL19,DBLP:conf/icdm/KoYBP0K21,DBLP:journals/tkde/LianXCX21,DBLP:conf/cikm/KangM19,DBLP:conf/sigir/ShiMZZYSLM20}.
Recently, discrete codes have gained popularity for improving recommendation quality in various scenarios.
They are particularly beneficial in alleviating data sparsity and offering prior semantics, which has proven advantageous in transferable recommendation~\cite{DBLP:conf/www/HouHMZ23}, generative sequential recommendation~\cite{DBLP:conf/nips/RajputMSKVHHT0S23,DBLP:journals/corr/abs-2309-13375,DBLP:journals/corr/abs-2405-07314,DBLP:journals/corr/abs-2405-16871} and LLM-based recommendation~\cite{DBLP:conf/sigir-ap/HuaXGZ23,DBLP:journals/corr/abs-2311-09049}. 
Different from these studies, our work aims to employ discrete codes for virtual neighbor augmentation and semantic similarity sampling to enhance graph CL in CF.

\section{Conclusion}

In this paper, we proposed a novel framework to enhance graph CL by constructing reliable and
informative contrastive views that imply stronger collaborative information.
The core idea is to learn discrete codes associated with rich collaborative information for users and items to generate contrastive views.
Specifically, we present an end-to-end multi-level vector quantizer to map users and items into discrete codes.
These codes are used to enhance the neighborhood structure and semantic relevance of contrastive views.
Firstly, we generate dual augmented nodes with abundant neighborhood structures by replacing node neighbors with discrete codes or adding them as virtual neighbors relying on the observed interactions.
Secondly, we consider users/items with shared discrete codes as semantically relevant and select similar positive examples based on this semantic relevance.
Finally, we introduce a triple-view graph contrastive learning approach to align two augmented nodes and the sampled similar user/item.
Extensive experiments on four public datasets demonstrate the effectiveness of our proposed CoGCL. 
As future work, we attempt to improve the scalability of our framework to extend it to other recommendation scenarios, such as click-through rate prediction and sequential recommendation.

\balance

% \begin{acks}
% \end{acks}

\bibliographystyle{ACM-Reference-Format}
\bibliography{ref}

\clearpage

\appendix

\section{Time and Space Complexity}
\label{apd:complexity}
\subsection{Time Complexity}
We analyze the time complexity of the following procedures in our CoGCL:
(1) The neighbor information aggregation based on LightGCN consumes $\mathcal{O}(L \times |\mathcal{E}| \times d)$ time, where $L$ denotes the number of GNN layers, and $d$ is the dimension of user/item embeddings.
(2) The time consumption for user and item discrete code learning is $\mathcal{O}(B \times H \times K \times d)$, where $B$ is the batch size, $H$ denotes the number of code levels, and $K$ represents the size of codebook. 
Thanks to the benefits of RQ or PQ allowing for a vast expression space (\ie $K^H$) with minimal codes~\cite{vasuki2006review,DBLP:journals/sensors/ChenGW10,DBLP:journals/pami/JegouDS11,DBLP:journals/taslp/ZeghidourLOST22}, in real-world applications, $H$ and $K$ typically satisfy $H*K \ll |\mathcal{U}|$ and $H*K \ll |\mathcal{I}|$~(\eg 4*256).
(3) To obtain contrastive view representations, it takes $\mathcal{O}(L \times (|\mathcal{E}^{o_1}| + |\mathcal{E}^{o_2}|) \times d)$ time to encode node representations based on the augmented graphs.
After training, only the time taken by the first part is retained for future recommendations, which is the same as LightGCN.

\subsection{Space Complexity}
Regarding space complexity, our CoGCL only introduces $\mathcal{O}(H \times K \times d)$ additional embedding parameters for discrete codes compared to LightGCN.
Also benefiting from the advantages of RQ or PQ in expression space, the value of $H*K$ is typically much smaller than the number of users and items~\cite{DBLP:conf/www/HouHMZ23,DBLP:conf/nips/RajputMSKVHHT0S23,DBLP:journals/corr/abs-2311-09049}.

\section{Supplement for Experiment}

% \subsection{Dataset Statistics}
% \label{apd:data_statistics}

% \begin{table}[H]
% \centering
% \caption{Statistics of the preprocessed datasets.}
% \label{tab:data_statistics}
% \begin{tabular}{lrrrr}
% \toprule
%  \textbf{Datasets} & \textbf{\#Users} & \textbf{\#Items}  & \textbf{\#Interactions} & \textbf{Sparsity} \\
% \midrule
% Instrument & 48,453& 21,413  & 427,674  & 99.959\% \\
% Office & 181,878 & 67,409 & 1,477,820   & 99.988\%  \\
% Gowalla & 29,858 & 40,988 & 1,027,464 & 99.916\%  \\ 
% iFashion  & 300,000 & 81,614 & 1,607,813 & 99.993\%  \\ 
% \bottomrule
% \end{tabular}
% \end{table}

\subsection{Baseline Models}
\label{apd:baseline}
We adopt the following competitive baselines for comparison with our CoGCL:

\noindent (1) \textbf{Traditional CF Models:}
\begin{itemize}
    \item {\textbf{BPR}}~\cite{DBLP:conf/uai/RendleFGS09} is a matrix factorization (MF) model to learn latent representations for users and items by optimizing the BPR loss.
    \item {\textbf{GCMC}}~\cite{DBLP:journals/corr/BergKW17} proposes a bipartite graph-based auto-encoder framework for matrix completion.
    \item {\textbf{NGCF}}~\cite{DBLP:conf/sigir/Wang0WFC19} adopts graph convolution for high-order relation modeling to improve the performance of recommendation.
    \item {\textbf{DGCF}}~\cite{DBLP:conf/sigir/WangJZ0XC20} learns disentangled representations for users and items to distill intent information.
    \item {\textbf{LightGCN}}~\cite{DBLP:conf/sigir/0001DWLZ020} simplifies GCN by removing feature transformation and nonlinear activation to make it more suitable for recommendation.
    \item {\textbf{SimpleX}}~\cite{DBLP:conf/cikm/MaoZWDDXH21} is a simple and strong baseline for collaborative filtering via cosine contrastive loss.
\end{itemize}

\noindent (2) \textbf{CL-based Models:}
\begin{itemize}
    \item {\textbf{SLRec}}~\cite{DBLP:conf/cikm/YaoYCYCMHCTKE21} uses contrastive learning for representation regularization to learn better latent relationships.
    \item {\textbf{SGL}}~\cite{DBLP:conf/sigir/WuWF0CLX21} introduces self-supervised learning to improve graph collaborative filtering. We adopt SGL-ED in our experiments.
    \item {\textbf{NCL}}~\cite{DBLP:conf/www/LinTHZ22} utilizes neighborhood-enriched contrastive learning to enhance GNN-based recommendation.
    \item {\textbf{HCCF}}~\cite{DBLP:conf/sigir/XiaHXZYH22} constructs hypergraph-enhanced contrastive learning to capture local and global collaborative relations.
    \item {\textbf{GFormer}}~\cite{DBLP:conf/sigir/LiXRY0023} leverages graph transformer to distill self-supervised signals with invariant collaborative rationales.
    \item {\textbf{SimGCL}}~\cite{DBLP:conf/sigir/YuY00CN22} creates contrastive views by adding random noise to the embedding space for graph contrastive learning.
    \item {\textbf{LightGCL}}~\cite{DBLP:conf/iclr/Cai0XR23}  employs singular value decomposition~(SVD) to generate augmented view for lightweight graph contrastive learning.
\end{itemize}

\subsection{Hyperparameter Tuning}
\label{apd:hyper_tune}
\begin{figure}[H]
	\centering
	\subfloat[$\mu$]{
            \includegraphics[width=1\linewidth]{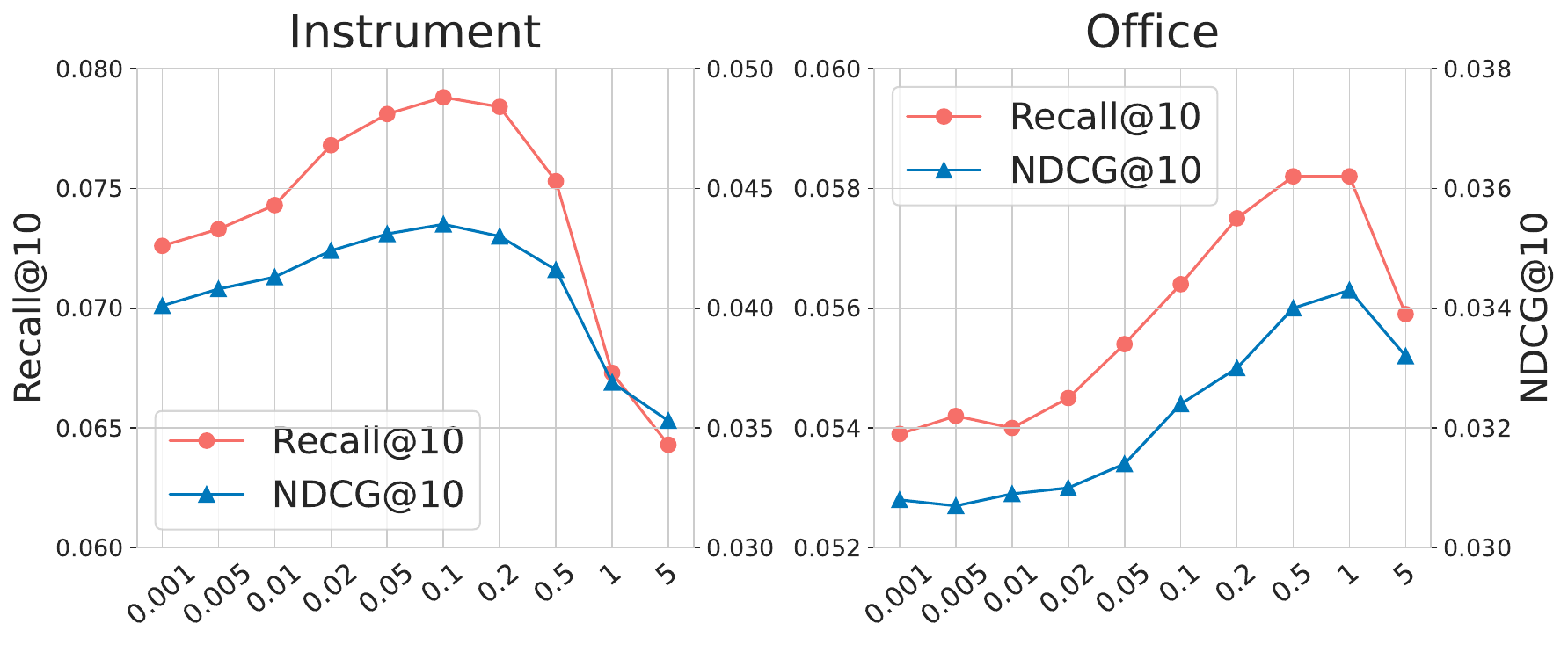}
        } \\
        \subfloat[$\eta$]{
            \includegraphics[width=1\linewidth]{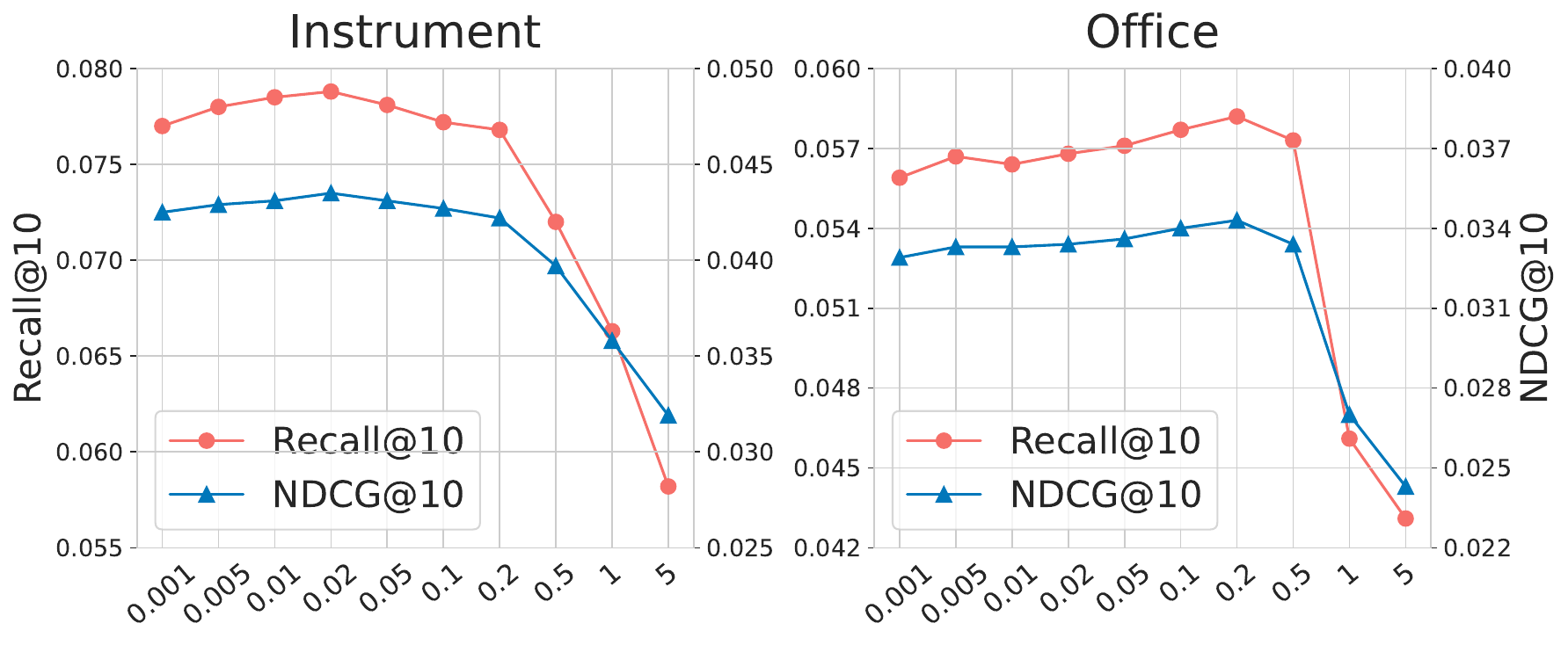}
        }
	\caption{Performance comparison of different CL loss coefficients.}
	\label{fig:cl_weight}
\end{figure}

We investigate the impact of the following hyperparameters on model performance:

\paratitle{CL loss coefficients $\mu$ and $\eta$}
% In this part, we investigate the sensitivity of our model to the CL loss coefficients $\mu$ and $\eta$ in Eq.~\eqref{eq:loss}. 
% Specifically, w
We tune $\mu$ and $\eta$ in the range of \{5, 1, 0.5, 0.2, 0.1, 0.05, 0.02, 0.01, 0.005, 0.001\}, the results are shown in Figure~\ref{fig:cl_weight}.

From the results, we can observe that too large or too small $\mu$ would lead to suboptimal performance, and too large $\eta$ would cause a sharp drop in performance.
The recommended values for $\mu$ on Instrument and Office datasets are 0.1 and 1, respectively, while the optimal values of $\eta$ for these datasets are 0.02 and 0.2, respectively.
Generally, the optimal value of $\eta$ is smaller than that of $\mu$, and it is essential to tune these hyperparameters for the balance between different objectives.

% \paratitle{Impact of code loss coefficient $\lambda$.}
% We tune $\lambda$ in range of \{0.1, 0.2, 0.5, 1, 2, 5, 10\}.
% As shown in Figure~\ref{fig:lambda}, the optimal values of $\lambda$ on Instrument and Office datasets are 5 and 1, respectively. In addition, the performance remains superior when the value is between $[0.1,5]$.

\paratitle{Augmentation probabilities.}

\begin{figure}[H]
	\centering
	\subfloat[Probability of ``\emph{replace}'']{
            \includegraphics[width=1\linewidth]{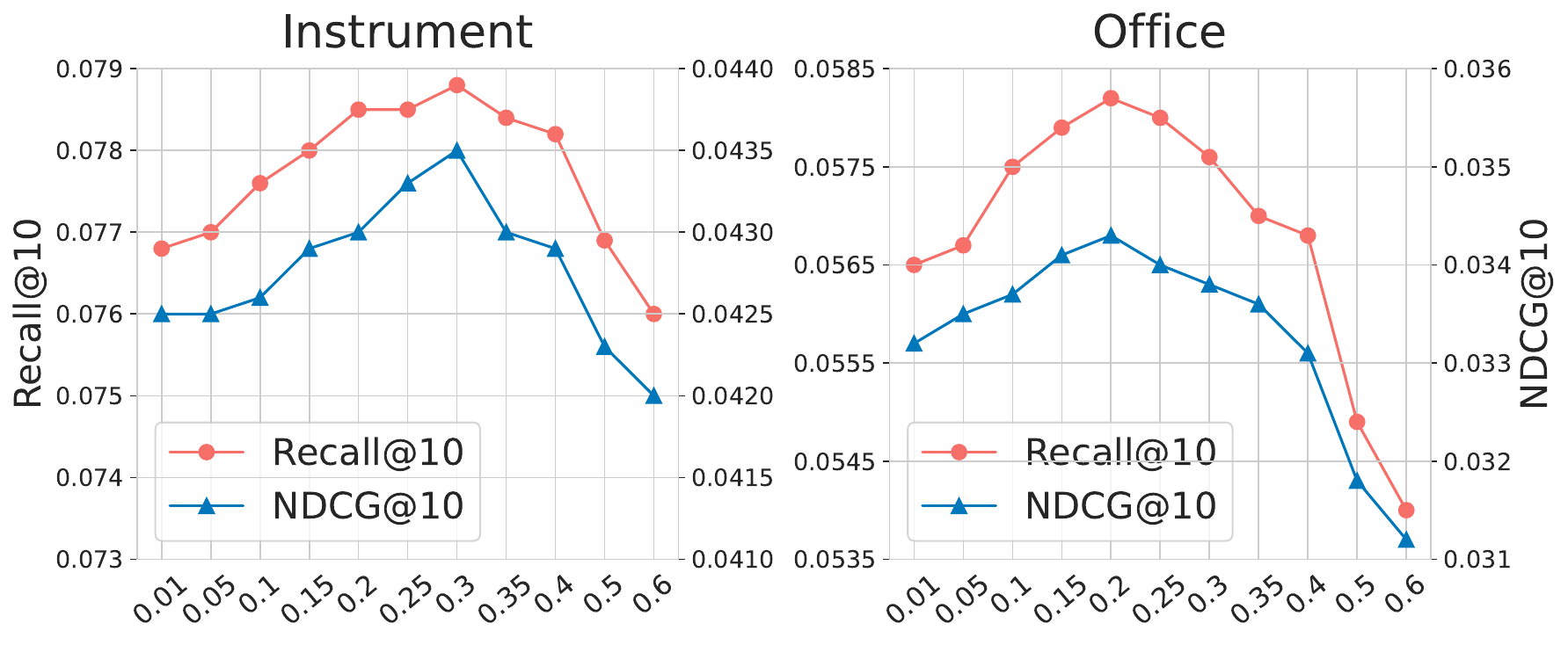}
        } \\
        \subfloat[Probability of ``\emph{add}'']{
            \includegraphics[width=1\linewidth]{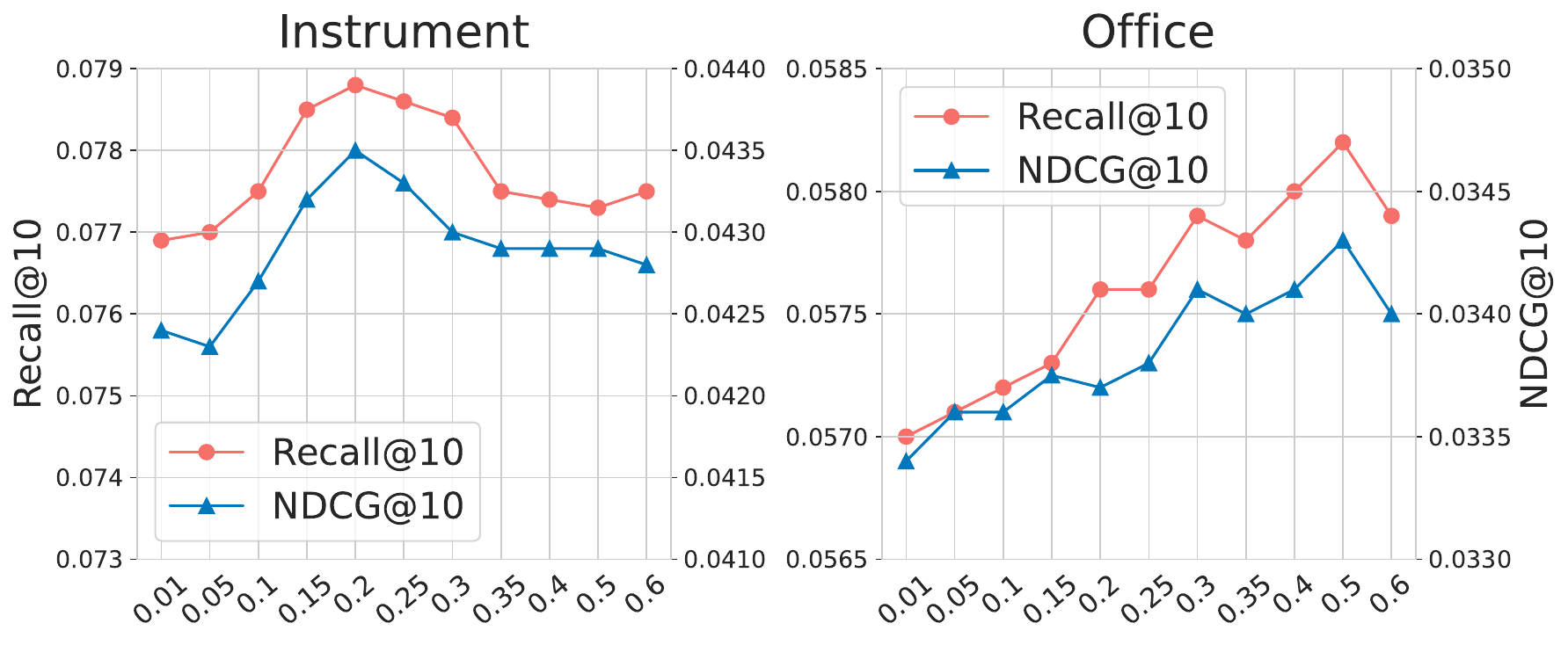}
        }
	\caption{Performance comparison of different augmentation probabilities.}
	\label{fig:aug_pro}
\end{figure}

We tune the probability of ``\emph{replace}'' and ``\emph{add}'' augmentation in the range of \{0.01, 0.05, 0.1, 0.15, 0.2, 0.25, 0.3, 0.4, 0.5, 0.6\}.
The results are shown in Figure~\ref{fig:aug_pro}.

We find that the probability of either ``\emph{replace}'' or ``\emph{add}'' should not be excessively high or too low.
In Instrument dataset, the ideal probability for ``\emph{replace}'' is 0.3 and for ``\emph{add}'' is 0.2.
In Office dataset, the optimal probability for ``\emph{replace}'' is 0.2 and for ``\emph{add}'' is 0.5.

\begin{figure}[H]
\centering
\includegraphics[width=1.0\linewidth]{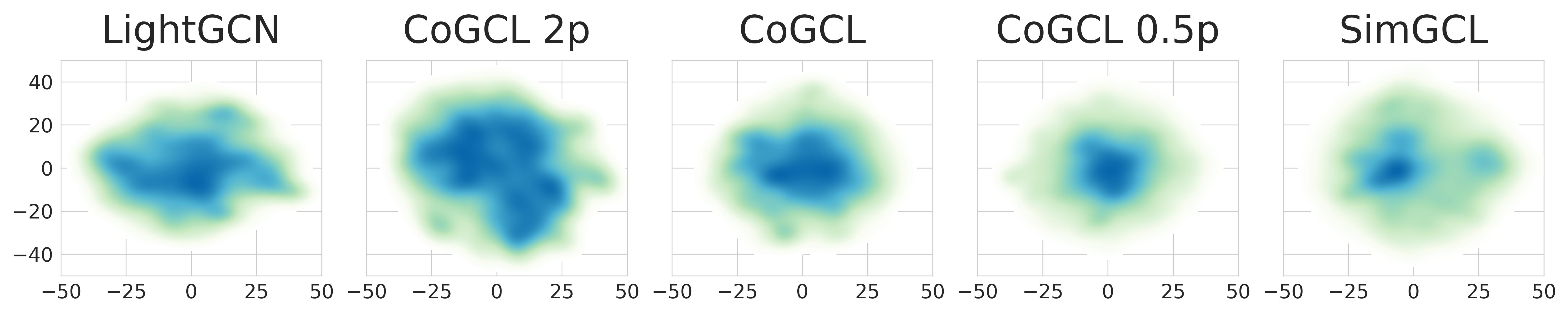}
\caption{Embedding distribution of different data augmentation ratios on Instrument dataset. The transition from green to blue signifies a gradual increase in embedding density.}
\label{fig:emb_distr}
\end{figure}

\subsection{Embedding Distribution \wrt Augmentation Ratio}
To more intuitively understand the contribution of CoGCL, we visualize the learned embedding distribution under different data augmentation ratios in Figure~\ref{fig:emb_distr}.
We first map user embeddings to two-dimensional space based on t-SNE~\cite{van2008visualizing}. 
Then we apply Gaussian kernel density estimation~(KDE)~\cite{botev2010kernel} to plot the user embedding distribution in the two-dimensional space.
\underline{w $2p$} and \underline{w $0.5p$} indicate that probabilities (both ``\emph{replace}'' and ``\emph{add}'') for virtual neighbor augmentation are adjusted to twice and half of the optimal values respectively.
From the results, we can find that the embedding distributions learned by CoGCL and SimGCL are more uniform than that of LightGCN, thanks to the uniformity brought by CL. 
Compared with SimGCL, the embedding learned by CoGCL achieves a good trade-off between clustering and uniformity.
In addition, it can be seen that the embeddings exhibit a more clustered pattern as the augmentation ratio rises, suggesting that higher augmentation probabilities lead to a stronger tendency for clustering.

\end{document}